\documentclass[times,twocolumn,review,nopreprintline]{elsarticle}

\usepackage{times}  
\usepackage{helvet}  
\usepackage{courier}  
\usepackage{url}  
\usepackage{graphicx}  
\usepackage{subcaption}
\usepackage{multirow}
\usepackage{amsmath}
\usepackage{mathrsfs}
\usepackage{algorithm}
\usepackage{algorithmic}
\usepackage{color}
\usepackage{comment}
\usepackage{amssymb}
\usepackage{amsmath}

\usepackage{booktabs,caption}
\usepackage{medima}
\usepackage{framed,multirow}

\newcommand{\revisedMR}[1]{{\color[rgb]{0.5,0.2,0.7}{#1}}}
\newcommand{\revised}[1]{{\color[rgb]{0.0,0.0,0.0}{#1}}}

\journal{Medical Image Analysis}

\begin{document}
\verso{Cheng Xue \textit{et~al.}}

\begin{frontmatter}

\title{Global Guidance Network for Breast Lesion Segmentation in Ultrasound Images }%

\author[1]{Cheng \snm{Xue}}
\author[2]{Lei \snm{Zhu}\corref{cor1}}
\cortext[cor1]{Corresponding author: Lei Zhu (lz437@cam.ac.uk)}
\author[3]{Huazhu \snm{Fu}}
\author[1]{Xiaowei \snm{Hu}}
\author[1]{Xiaomeng \snm{Li}}
\author[4]{Hai \snm{Zhang}}
\author[1]{Pheng-Ann \snm{Heng}}


\address[1]{Department of Computer Science and Engineering, The Chinese University of Hong Kong}
\address[2]{Department of Applied Mathematics and Theoretical Physics, University of Cambridge}
\address[3]{Inception Institute of Artificial Intelligence, Abu Dhabi, UAE}
\address[4]{Shenzhen People's Hospital, The Second Clinical College of Jinan University}

\begin{abstract}
Automatic breast lesion segmentation in ultrasound helps to diagnose breast cancer, which is one of the dreadful diseases that affect women globally. Segmenting breast regions accurately from ultrasound image is a challenging task due to the inherent speckle artifacts, blurry breast lesion boundaries, and inhomogeneous intensity distributions inside the breast lesion regions. Recently, convolutional neural networks (CNNs) have demonstrated remarkable results in medical image segmentation tasks. However, the convolutional operations in a CNN often focus on local regions, which suffer from limited capabilities in capturing long-range dependencies of the input ultrasound image, resulting in degraded breast lesion segmentation accuracy. \revised{In this paper, we develop a deep convolutional neural network equipped with a global guidance block (GGB) and breast lesion boundary detection (BD) modules for boosting the breast ultrasound lesion segmentation. The GGB utilizes the multi-layer integrated feature map as a guidance information to learn the long-range non-local dependencies from both spatial and channel domains.
The BD modules learn additional breast lesion boundary map to enhance the boundary quality of a segmentation result refinement.}
Experimental results on a public dataset and a collected dataset show that our network outperforms other medical image segmentation methods and the recent semantic segmentation methods  on breast ultrasound lesion segmentation.
Moreover, we also show the application of our network on the ultrasound prostate segmentation, in which our method better identifies prostate regions than state-of-the-art networks.
\end{abstract}
\end{frontmatter}


\section{Introduction}
\begin{figure}[t!]
	\centering
	\includegraphics[width=0.48\textwidth]{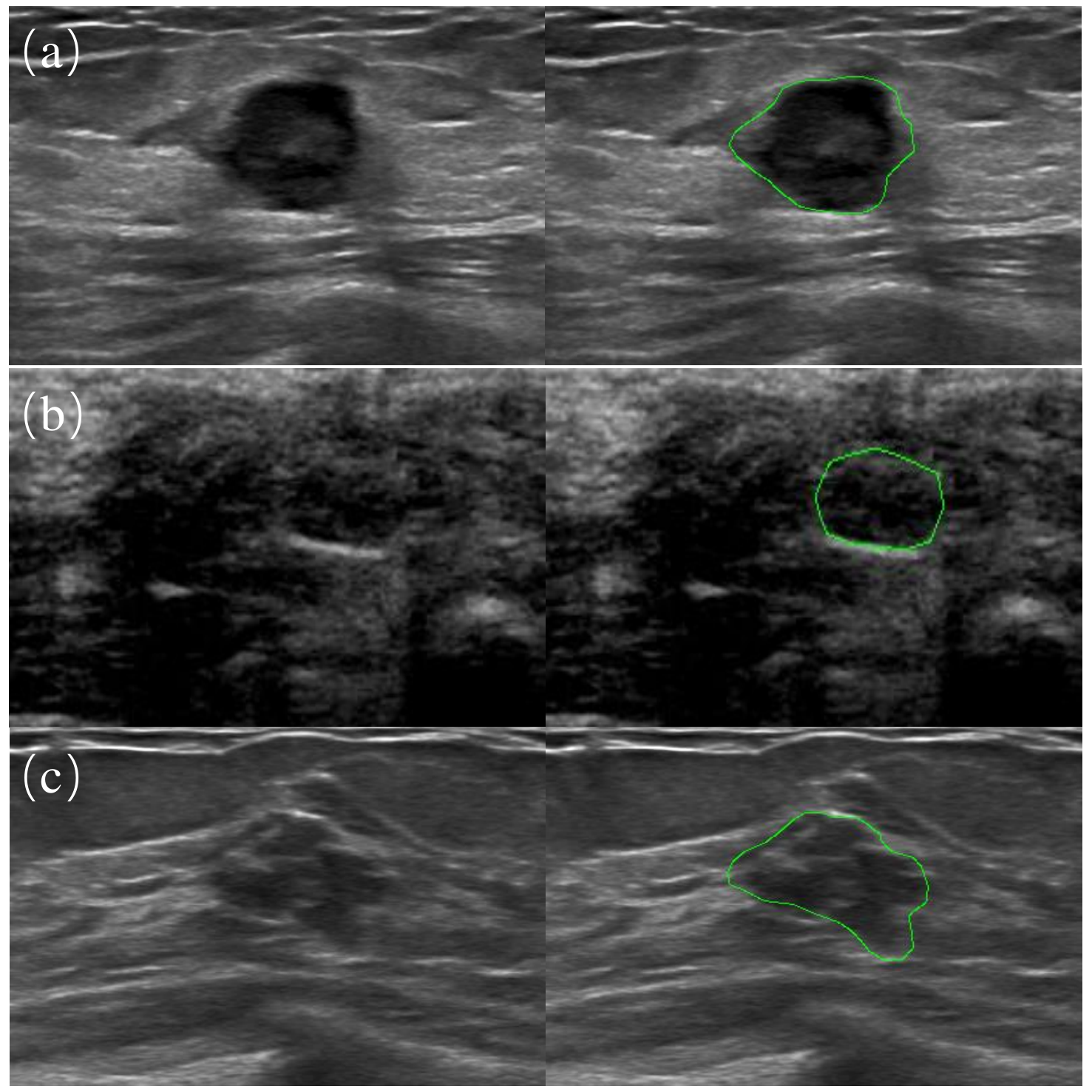}
	\caption{Examples of challenging cases in breast ultrasound lesion segmentation. The green contour denotes the breast lesion boundary. Left: the input ultrasound images. Right: the lesion region. (a) Inhomogeneous distributions inside the breast lesion region. (b) Ambiguous boundary due to similar appearance \revisedMR{between lesion regions and non-lesion backgrounds}. (c) Irregular breast lesion shapes.}
	\label{fig:1}
\end{figure}

Breast cancer is one of the dreadful diseases that affect women globally.
According to the statistic information reported in~\citep{American2019}, an estimated 42,260 breast cancer deaths \revisedMR{would occur in 2019}.
An accurate breast lesion segmentation from the ultrasound images helps the early diagnosis of breast cancer.
However, the automatic breast lesion segmentation in a 2D ultrasound image is a challenging task, since there are the speckle noise, and strong shadows in the ultrasound, inhomogeneous distributions in the breast lesion regions, and ambiguous boundaries between the breast lesion and non-lesion regions, as well as the irregular breast lesion shapes; see Fig.~\ref{fig:1} for the examples.

Segmenting breast lesion in ultrasound images has been widely studied in the research community.
Early attempts, e.g.,~\citep{shan2012completely,madabhushi2002automatic,shan2008novel,kwak2005rd,madabhushi2003combining,yezzi1997geometric,chen2002cell,xian2015fully,ashton1995multiple,boukerroui1998multiresolution,xiao2002segmentation} detected the breast lesion boundaries mainly based on the hand-crafted features. These features, however, have the limited feature representation ability, leading to misrecognize the breast lesions in a complex environment.
Recently, the convolutional neural networks (CNNs) have achieved impressive progress on breast ultrasound segmentation task.
\revised{For examples, Yap \textit{et al.}, adopted U-Net, FCN-AlexNet, and patch-based LeNet for 2D ultrasound image breast lesion detection~\citep{yap2017automated}}.  Lei \textit{et al.}, employed a deep neural network with the supervision signals on the boundary to address the whole breast ultrasound image~\citep{lei2018segmentation}.
 Xu \textit{et al.},  adopted an eight-layer CNN to segment 3D breast in the ultrasound data~\citep{xu2019medical}.

The ultrasound image has many distant pixels, which have the similar appearance as the breast lesions.
Incorporating these pixels could provide long-term non-local features to learning discriminative features for the ultrasound breast lesion segmentation.
Capturing the global contextual information for ultrasound image segmentation is a long-standing topic in the medical image community. Previous studies proposed to enlarge the receptive field with dilated convolutions, pooling operations~\citep{chen2018encoder,chen2017deeplab,chen2014semantic}; or fuse the middle level and high level features with more task-related semantic features~\citep{ronneberger2015u,lin2017feature}.
However, these methods fail to capture the contextual information in a global view and only consider the inter-dependencies among spatial domains.
In medical image analysis community, most previous approaches rely on local region operation for segmentation task~\citep{ronneberger2015u,dou20163d,lin2017feature}.
However, capturing the long-range dependencies information holds promising potentials but has not been well explored yet.
\revised{
Traditional non-local blocks in these networks~\citep{qi2019x,dou2018local} are only embedded into the deep CNN layers to learn long-range dependencies for network predictions. However, due to the relatively larger receptive fields than shallow CNN layers, the deep layers of a segmentation network are responsible for capturing cues of the whole breast lesions and somehow lack parts of breast lesion regions, degrading the segmentation performance.

In this work, we develop a convolutional neural network (CNN) to integrate features at all CNN layers (including deep and shallow CNN layers) to produce multi-level integrated features (MLIF) as a guidance information of the non-local blocks in spatial and channel manners to complement more breast lesion boundary details, which are usually neglected by deep CNN layers.
Moreover, we propose to predict additional breast lesion boundary map such that the predicted boundary map is regularized to be as similar as the underlying ground truth.
By doing so, our network can produce a segmentation result with more accurate breast lesion boundaries.
}
In summary, our contributions are four-fold:
\begin{itemize}
	\item First, \revised{we present a CNN (denoted as GG-Net) with a global guidance block (GGB)  to aggregate non-local features in both spatial and channel domains under the guidance of multi-layer integrated features for learning a powerful non-local contextual information.}
	\item Second, \revised{we develop a breast lesion boundary detection (BD) module in shallow CNN layers to embed additional boundary maps of breast lesions for obtaining the segmentation result with high-quality boundaries.}
	\item Third, the experimental results on two ultrasound breast lesion datasets show that our network outperforms the state-of-the-art medical image segmentation methods on breast lesion segmentation.
    \item Moreover, we also show the application of our network on the ultrasound prostate segmentation, where our network obtains satisfactory performance.
\end{itemize}

\section{Related works}

Breast lesion segmentation from ultrasound images is very challenging due to the speckle artifacts, low contrast, shadows, blurry boundaries, and the variance in lesion shapes~\citep{kirberger1995imaging}.
A variety of breast lesion segmentation algorithms have been proposed and these methods can be broadly classified into four categories, including region based approach~\citep{shan2012completely,madabhushi2002automatic,shan2008novel,kwak2005rd}, deformable models~\citep{madabhushi2003combining,yezzi1997geometric,chen2002cell}, graph-based approaches~\citep{xian2015fully,ashton1995multiple,boukerroui1998multiresolution,xiao2002segmentation} and learning based approaches~\citep{liu2010fully,huang2008neural,lo2014computer,moon2014tumor,othman2011segmentation}.
These approaches usually employed texture features to represent the local variation of pixel intensities and then detect abnormal regions in the ultrasound image.
However, these methods rely on hand-crafted features and have limited representation capacity.

Convolutional neural networks (CNNs) have shown remarkable performance in many medical image analysis tasks, including image classification~\citep{yu2018melanoma,yu2017deep}, semantic segmentation~\citep{ronneberger2015u,dou20163d,yu2016automated,li2018dense}.
These methods utilized the superior learning capability of neural network and outperformed other traditional segmentation methods.
\revised{For breast image analysis, recent works have featured CNN based methods~\citep{yap2017automated,lei2018segmentation,xu2019medical,dhungel2017deep,mordang2016reducing,ahn2017novel,mordang2016automatic,hu2019automatic,mishra2018ultrasound}. Yap \textit{et al.} adopted pacth-based LeNet, U-Net, and FCN-AlexNet for breast lesion detection}~\cite{yap2017automated}. Lei\textit{et al.} proposed a ConvEDNet for whole breast ultrasound image segmentation with the deep boundary supervision and adaptive domain transfer knowledge~\citep{lei2018segmentation}.
Some works adopted CNNs with different layers to detect mass, estimate the breast density, and segment breast ultrasound images~\citep{dhungel2017deep,ahn2017novel,xu2019medical}. Mordang \textit{et al.}, adopted OxfordNet for mammography microcalcification detection~\citep{mordang2016automatic}.
Hu \textit{et al.}, proposed a dilated fully convolutional network for breast tumor segmentation~\citep{hu2019automatic}. Mishra \textit{et al.}, developed a fully convolutional neural network with deep supervision for \revisedMR{lumen segmentation and liver lesion segmentation}~\citep{mishra2018ultrasound}.

\begin{figure*}[t!]
	\centering
	\includegraphics[width=1\textwidth]{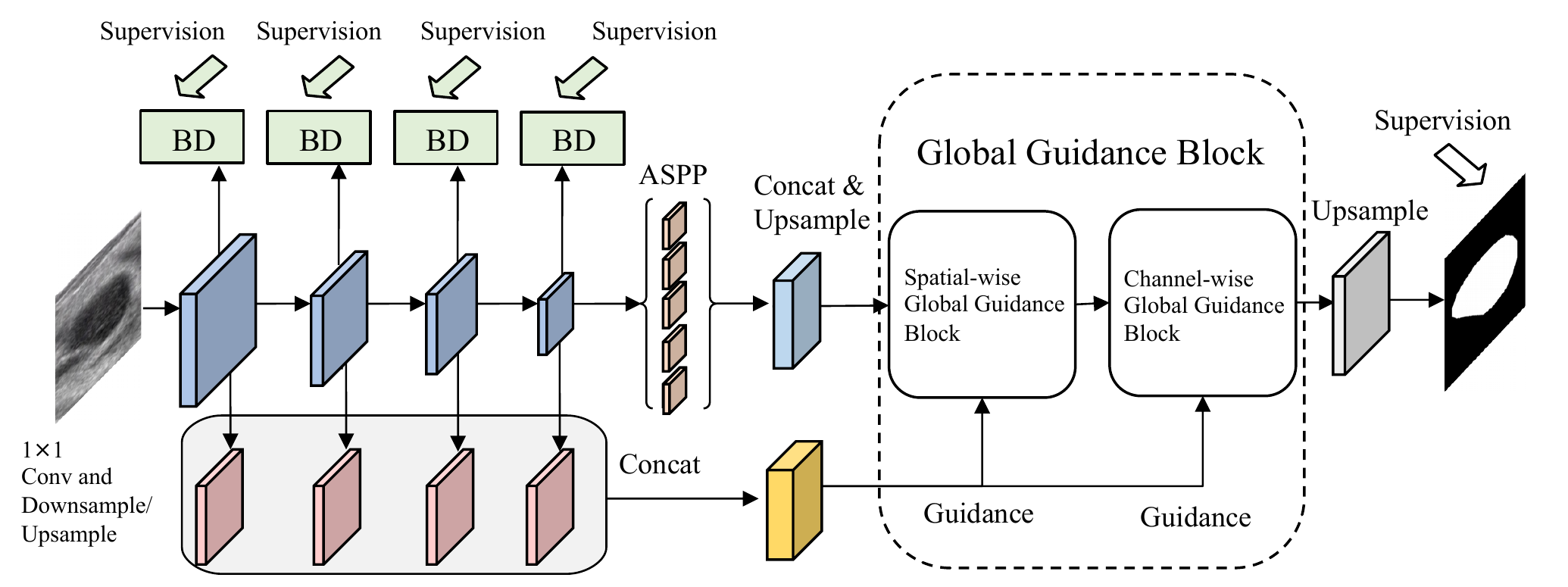}
	\caption{\revised{The schematic illustration of the proposed breast lesion segmentation network (GG-Net) in this work. (i) We first use a convolutional neural network (CNN) to produce a set of feature maps with different scales, followed by a ASPP module to enlarge the receptive field. (ii) In each CNN layer, we pass its feature map to a breast lesion boundary detection (BD) module (see Section III. B) to detect breast lesion boundaries. (iii) We concatenate features at all CNN layers and use it as the guidance to the developed global guidance block (GGB), which includes a spatial-wise global guidance block and a channel-wise global guidance block, to learn long-range dependencies for each pair of positions on the feature maps over spatial and channel domains. (iv) We use the output feature map of the GGB to predict the segmentation result of our network.}}
	\label{fig:2}
\end{figure*}

To improve the pixel-wise prediction accuracy, many researchers considered incorporating the long-range dependencies and contextual information in the network, thus enhancing the feature representation for pixel-wise prediction.
\revisedMR{For example}, atrous spatial pyramid pooling (ASPP) was designed to embed the global contextual information, and it was widely adopted in DeepLabv2~\citep{chen2014semantic} and DeepLabv3~\citep{chen2018encoder}.
Similarly, Zhao \textit{et al.}, designed a pyramid pooling module to collect the effective contextual prior with different scales~\citep{zhao2017pyramid}. Besides, an EncNet was  introduced a channel attention mechanism to capture the global context~\citep{zhang2018context}. Peng \textit{et al.}, argued that large kernel plays an important role in semantic segmentation tasks, and \revisedMR{a global convolutional network} was proposed to learn the context information~\citep{peng2017large}.
In medical image analysis field, there are some recently work that also considered the context information, such as the encoder-decoder structures~\citep{ronneberger2015u} fused the mid-level and high-level features to obtain different scale context.
In OBELISK-Net~\citep{heinrich2019obelisk}, sparse deformable convolutions were formulated to learn large context information.
However, these methods mostly stacked a series of convolutional layers to capture the context information.
Several works have been proposed to alleviate this issue by implicitly utilizing attention mechanisms or non-local operations to increase the receptive fields and capture contextual information~\citep{wang2018non,vaswani2017attention,schlemper2019attention,zhang2017mdnet,roy2018recalibrating,joutard2019permutohedral}.
However, the meticulous features in the multi-layer features and the long range dependencies between feature channels are ignored.
In this regard, we introduce a network that gracefully unifies the approaches mentioned above, which not only consider the long-range dependencies spatial-wisely and channel-wisely, but also embed contextual information from different layers.

\section{Methodology}

Fig.~\ref{fig:2} illustrates the architecture of the developed network (denoted as \revised{GG-Net}).
Our network takes a breast ultrasound image as the input and produces a segmented  mask in an end-to-end manner.
Specifically, our GG-Net starts by using a CNN to generate multi-level feature maps with different spatial resolutions and adopting the ASPP~\citep{chen2018encoder} to enhance the receptive field of features.
In order to utilize the complementary information among different CNN layers, the \revised{GGB} is introduced to refine the features by learning long-range feature dependencies under the guidance of an integrated feature map from the shallow CNN layers.
Moreover,  the \revised{BD} module is embedded in the shallow CNN layers to capture the breast lesion contour and provide a strong cue for better segmenting breast lesions and refining lesion boundaries.
Finally, the prediction map is produced as the segmentation result of our network.
In the following subsections, we will introduce details of the developed GGB and BD in our method.

\subsection{\revised{Global Guidance Block}}
\begin{figure*}[t!]
	\centering
	\includegraphics[width=0.95\textwidth]{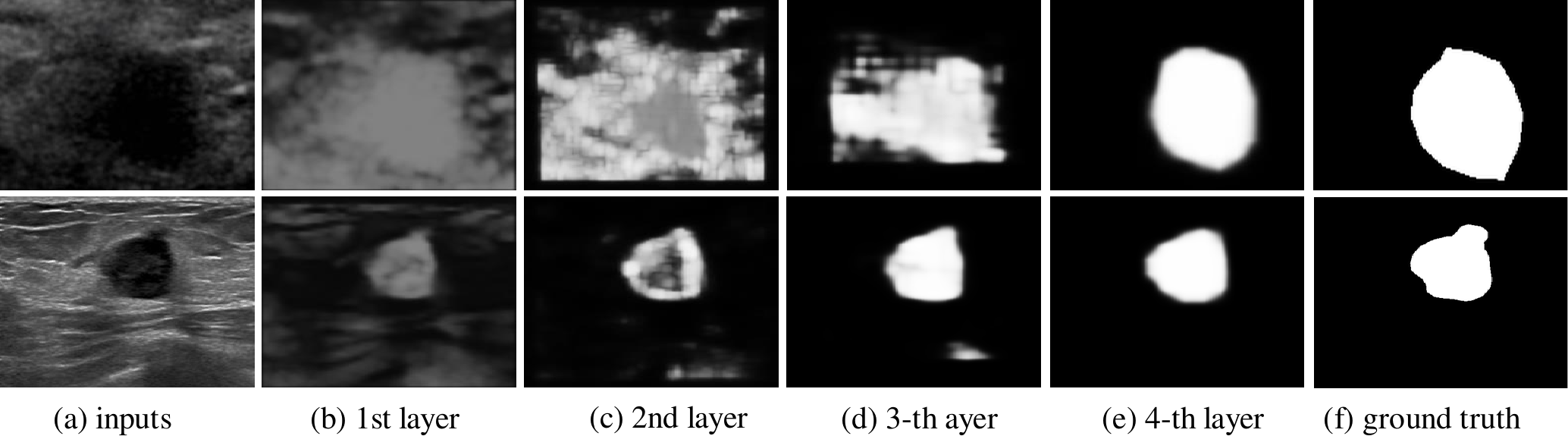}
	\caption{ Two examples are shown to illustrate the learned breast lesion feature on different layers. (a) Input images. (b)-(e) Segmentation maps predicted from the feature map from the 1-st layer to the 4-th layer. (f) Ground truths. The shallow layers (b), (c) and (d) contains more detail features compared to (e).}
	\label{fig:layer}
\end{figure*}

Convolutional and recurrent operations of CNNs only capture the spatial dependencies within a local neighborhood.
Although stacking convolutional layers can learn the long range dependencies, such repeating local convolutions is time-consuming and leads to the optimization difficulties that need to be carefully addressed~\citep{wang2018non}. Moreover, breast ultrasound images usually contain speckles and shadows that tend to be recognized as breast lesion due to the limited receptive fields of local convolutions.
In this regard, we develop a global guidance block (GGB), which leverages a guidance feature map to learn the long range dependencies by considering spatial and channel information.

\begin{figure}[t!]
	\centering
	\includegraphics[width=0.49\textwidth]{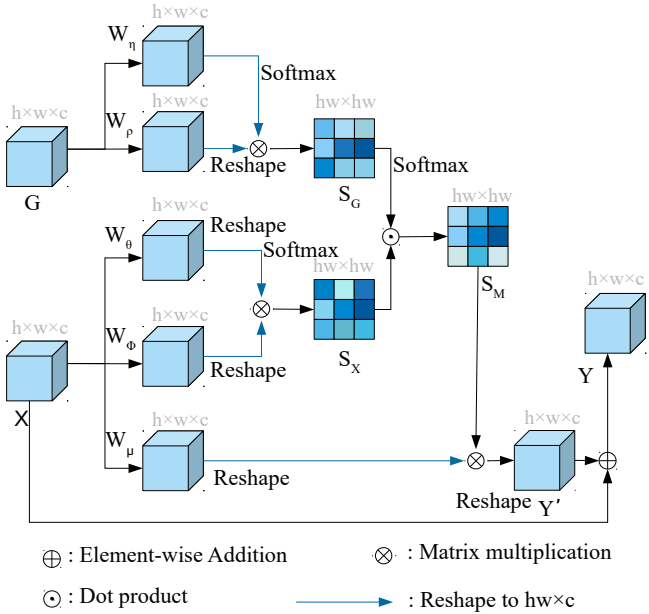}
	\caption{The schematic illustration of the details of spatial-wise GGB, where $G$ is the guidance map, and $X$ is the input feature map. }
	\label{fig:sgnlb}
\end{figure}

\subsubsection{\revised{Spatial-wise Global Guidance Block}}
\label{sec:sgnlb}
The feature maps from the shallow CNN layers provide detailed information but contain more non-lesion regions, while the deep CNN layers with larger reception fields eliminate the non-lesion regions, but tend to lose the local details.
In this regard, we argue that feature maps at different CNN layers contain the complementary information, as shown in Fig.~\ref{fig:layer}.
In our method, we first resize the feature maps of the first four CNN layers to the size of feature map from the second CNN layer, and then concatenate them to one multi-layer integrated feature (MLIF) map. After that, a \revised{spatial-wise global guidance block (spatial-wise GGB) is proposed to learn the long-range position dependencies by taking MLIF as a guidance map.}

Fig.~\ref{fig:sgnlb} shows the schematic illustration of our \revised{spatial-wise GGB}.
Specifically, let $X$ ($x \in \mathbb{R}^{h\times w \times c}$) denote the output feature map of the ASPP module (see Fig.~\ref{fig:2}), and $G$ ($g \in \mathbb{R}^{h\times w \times c}$) denotes the guidance map.
The \revised{spatial-wise GGB} first feeds $X$ into three $1\times 1$ convolution layers with different parameters, $W_{\theta(x)}$, $W_{\phi(x)}$, and $W_{\mu(x)}$), to generate three feature maps, $\theta(x)$, $\phi(x)$, and $\mu(x)$, respectively.
After that, we reshape $\theta(x)$, $\phi(x)$, and $\mu(x)$ as $\mathbb{R} ^{hw \times c}$ matrices,
multiply the reshaped $\phi(x)$ with the transpose of the reshaped $\theta(x)$, and apply a softmax layer on the multiplication result to compute a $hw \times hw$ spatial-wisely position similarity map $S_x$:
\begin{equation}
\label{equ:S_X}
S_x=\textit{Softmax}(X^T W^T_{\theta(x)} W_{\phi(x)} X) \ ,
\end{equation}
\revised{where $softmax$ follows the traditional sigmoid function and it is applied on each element of the $hw$ $\times$ $hw$ $X^T W^T_{\theta(x)} W_{\phi(x)} X$.} On the other side,  two $1\times 1$ convolution layers with parameters, $W_{\eta(g)}$, and $W_{\rho(g)}$), are applied on guidance map $G$ to obtain two feature maps, $\eta(x)$  and $\rho(x)$, reshape $\eta(x)$ and $\rho(x)$, multiply the reshaped $\eta(x)$ to the transpose of the reshaped $\rho(x)$, and apply a softmax layer for producing another $hw \times hw$  position similarity matrix (denoted as $S_g$) from the guidance map $G$:
\begin{equation}
\label{equ:S_G}
S_g=\textit{Softmax}(G^T W^T_{\rho(g)} W_{\eta(g)} G)) \ .
\end{equation}
Once obtaining two similarity matrices $S_X$ and $S_G$, we use a softmax layer on the element-wise multiplication result of $S_X$ and $S_G$ to generate a guided similarity matrix $S_M$.
Then, we multiply $S_M$ with the features $\mu(x)$ to obtain a new feature map $Y^{'}$, which is then added with the input features $X$ to generate the output feature map $Y$:
\begin{equation}
\label{equ:y}
Y= \mu(x) \ \textit{Softmax} (S_x \cdot S_g) + X \ .
\end{equation}

\subsubsection{\revised{Channel-wise Global Guidance Block}}
\label{sec:cgnlb}

\begin{figure}[t!]
	\centering
	\includegraphics[width=0.49\textwidth]{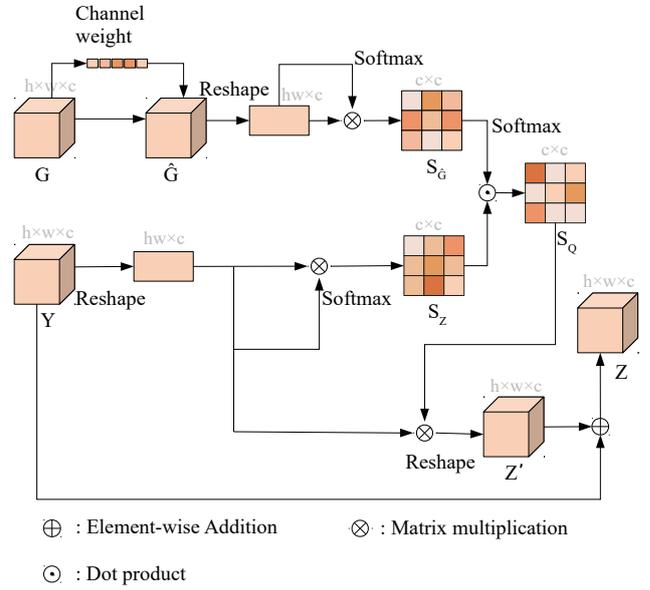}
	\caption{The schematic illustration of the channel-wise GGB, where $G$ is the guidance map and $Y$ is the input feature map. }
	\label{fig:cgnlb}
\end{figure}

Our \revised{spatial-wise GGB} treats each feature channel equally when learning the long range dependencies, resulting in neglecting the correlations among different feature channels.
Recently, allowing varied contributions from different feature channels has achieved superior performance in many computer vision tasks~\citep{hou2019interaction,chen2017sca,hu2018squeeze}.
Motivated by these, we develop a \revised{channel-wise global guidance block (channel-wise GGB)} to further learn the long range inter-dependencies between different feature channels.
Fig.~\ref{fig:cgnlb} illustrates the schematic details of the proposed \revised{channel-wise GGB}, which takes a feature map $Y$ and a guidance map $G$ as two inputs and generates a refined feature map  $Z$.
Specifically, we reshape $Y$ to $\mathbb{R}^{c \times hw}$,  multiply the reshaped $Y$ and the transpose of the reshaped $Y$, and use a softmax layer to obtain a channel-wise similarity map $S_Z \in \mathbb{R}^{c \times c}$.
Regarding the input guidance feature map $G$, we first use squeeze-and-excitation block to emphasis informative feature channels of $G$ and suppress less useful ones.
To achieve this, we use a global average pooling to generate the channel-wise statistics $\beta$, and the $k$-th element of the descriptor ($\beta$) is given by
\begin{equation}  \label{gate_step1}
\beta_k = \frac{1}{h\times w} \sum_{i=1}^{h}\sum_{j=1}^{w} G(i, j, k) \ ,
\end{equation}
where $G(i, j, k)$ denotes the element at the position $(i, j ,k)$ of guidance map $G$.
After that, we use two fully connected (fc) layers and a sigmoid activation function on the  channel-wise statistics $\beta$ to generate a coefficient vector $V_\lambda$:
\begin{equation}  \label{gate_step2}
V_\lambda = \Phi(W_2\Omega(W_1\beta)) \ ,
\end{equation}
where $W_1$ and $W_{2}$ denote the parameters of the two fully connected layers,
$\Omega$ and $\Phi$ are the ReLU and the sigmoid activation function, respectively.
Then, we multiply $V_\lambda$ with $G$ to assign different weights on channels of $G$ and obtain a refined feature map (denoted as $\hat{G}$).
Once obtaining $\hat{G}$, we reshape it to $\mathbb{R}^{c \times hw}$, multiple the reshaped $\hat{G}$ and the transpose of the reshaped $\hat{G}$, and use a softmax layer to generate a $c \times c$ similarity map $S_{\hat{G}}$.
Later, a softmax layer is applied on the multiplication of $S_Z$ and $S_{\hat{G}}$ to obtain a guided similarity map $S_Q$.
Finally, we multiply $S_Q$ with the input $Y$ to produce a new feature map $Z^{'}$, which is then added to the input features $Y$ to obtain the output feature map $Z$ of our channel-wise GGB.

\begin{figure}[t!]
	\centering
	\includegraphics[width=0.49\textwidth]{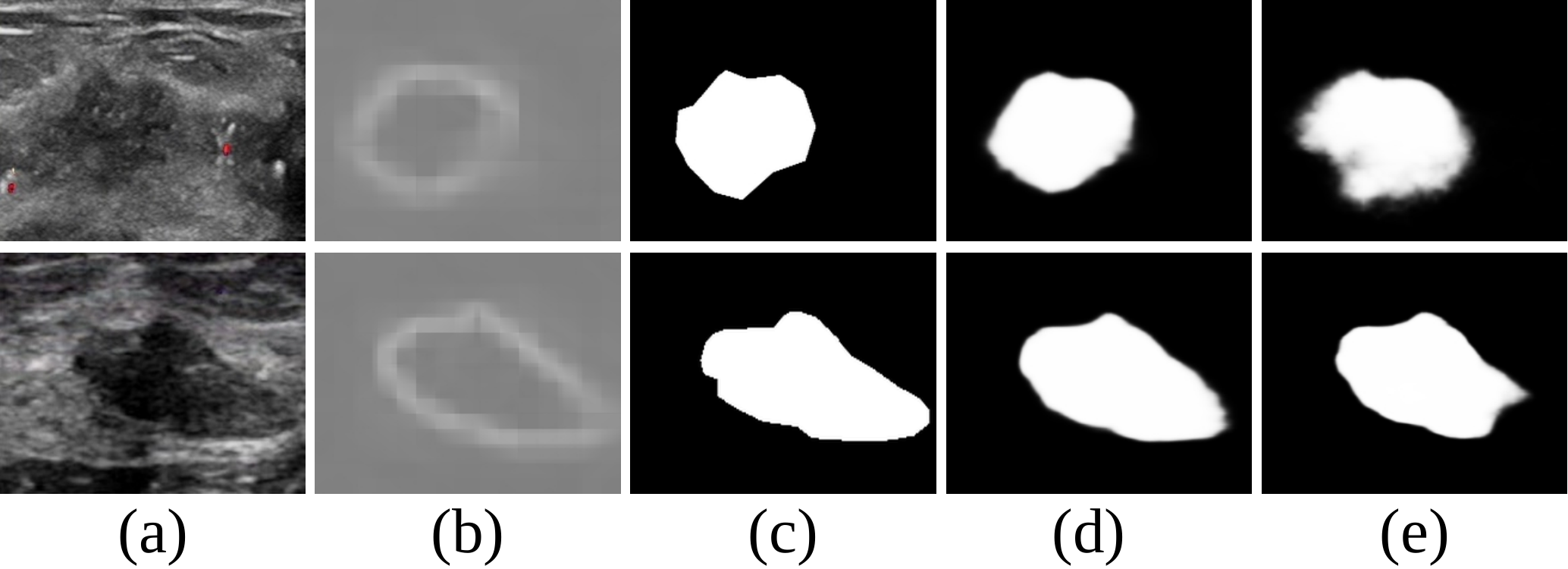}
	\caption{An analysis of segmentation improvement based on detected boundaries. (a) Input images. (b) Detected boundary map at \revised{BD} module at the fourth CNN layer. (c) Ground truths of breast lesion segmentation. (d) Segmentation results of our method. (e) Our results without the \revised{BD} module. Apparently, learning additional boundary maps of breast lesion incurs a better segmentation result.}
	\label{fig:bd}
\end{figure}

\subsection{Breast Lesion Boundary Detection Module}
\label{sec:boundary}

\begin{figure}[t!]
	\centering
	\includegraphics[width=0.5\textwidth]{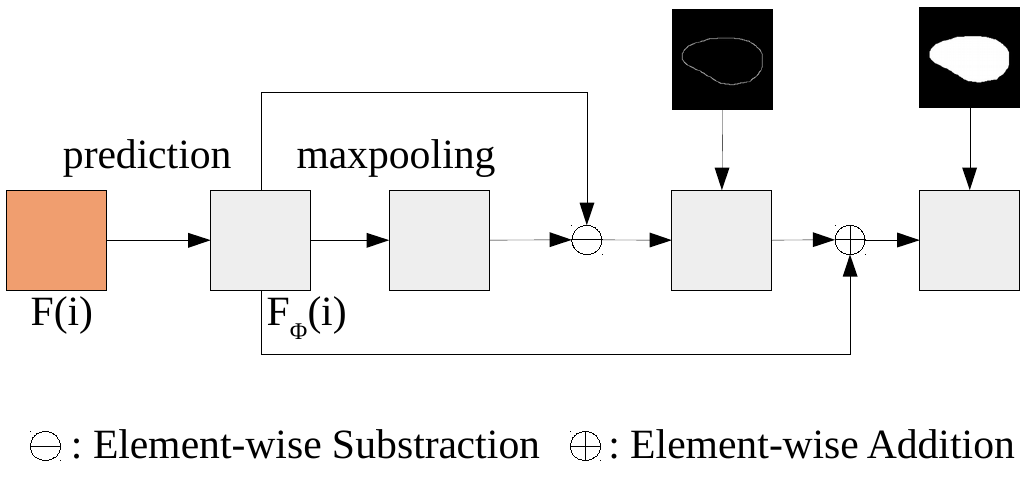}
	\caption{The schematic illustration of the breast lesion boundary detection (\revised{BD}) module. $F(i)$ is the feature map at the $i$-th CNN layer.}
	\vspace{-5mm}
	\label{fig:edge}
\end{figure}

Although \revised{GGB} generates a breast lesion segmentation result, we find that there are many failed segmented regions in the results, as shown in Fig.~\ref{fig:bd}(e), which have inaccurate boundary maps of the breast lesion.
To alleviate this, we develop a breast lesion boundary detection (\revised{BD}) module to identify multi-level boundary maps of the breast lesions and enhance the segmentation result with an additional boundary prediction loss.
Fig.~\ref{fig:edge} shows the schematic illustration of the developed BD module at the i-th CNN layer to detect breast lesion boundaries.
It takes the feature map of i-th CNN layer as the input and outputs a boundary map of the breast lesion and a breast lesion segmentation result.
Specifically, we first use a $1$$\times$$1$ convolutional layer on the input features $F(i)$ to obtain a new feature map $F_{\phi}(i)$ with one channel. Then, we shift $F_{\phi}(i)$ with one pixel via a maxpooling operation (stride = 1, padding = 1, kernel size = $3$$\times$$3$; see~\citep{feng2019attentive} for details) and subtract the shifted result from $F_{\phi}(i)$ to obtain a boundary map $E$ of the breast lesions.
After that, we add $F_{\phi}(i)$ with $E$ to obtain a breast lesion segmentation map.

\subsection{Loss Function}

As shown in Fig.~\ref{fig:2}, we add a \revised{BD} module
for the shallow CNN layer to jointly locate breast lesions and detect a boundary map from feature map at the CNN layer.
Hence, our network generates four boundary maps and four breast lesion segmentation results at four CNN layers.
Moreover, our network generates a final segmentation result of breast lesions from the \revised{GGB}.
With an annotated breast lesion mask, we apply a canny operator~\citep{canny1986computational}  to obtain the boundary mask as the ground truth of the boundary prediction. Finally, we compute the total loss of our network as:
\begin{equation}
\label{eq:total-loss}
L_{total}=\sum_{i=1}^{N_{layer}} (\lambda_1 \cdot {L^i}_{seg} + \lambda_2 \cdot {L^i}_{boundary} ) + {L^{f}}_{seg} \ ,
\end{equation}
where $N_{layer}$ is the number of CNN layers, and we empirically set $N_{layer}$ as four in our implementation.
${L^i}_{seg}$ and ${L^i}_{boundary}$ denote the segmentation loss  and the boundary loss in the BLBD module of $i$-th CNN layer, respectively. ${L^{f}}_{seg}$ is the loss function of the final segmentation result.
The weights $\lambda_1$ and $\lambda_2$ are to balance ${L^i}_{seg}$, ${L^i}_{boundary}$, and ${L^{f}}_{seg}$, and their values are empirically set as $\lambda_1=1$ and $\lambda_2 = 10$.

Let $\mathcal{P}^i$ denote the predicted breast lesion segmentation result at $i$-th CNN layer and $\mathcal{G}$ is the ground truth of the annotated breast lesion mask. ${L^i}_{seg}$ combines a dice coefficient loss and a binary cross-entropy loss to compute the difference between $\mathcal{P}^i$ and $\mathcal{G}$:
\begin{equation}
\label{eq:loss-i-seg}
{L^i}_{seg}= 1- \dfrac{2\sum_{j=1}^{N_p} (\mathcal{P}^i)_j \times (\mathcal{G})_j}{\sum_{j=1}^{N_p} (\mathcal{P}^{i})_j^2+\sum_{j=1}^{N_p} (\mathcal{G})_j^2} -\dfrac{1}{N_p}\sum_{j=1}^{N_p} (\mathcal{P}^i)_j log (\mathcal{G})_j \ ,
\end{equation}
where $N_p$ is the number of pixels in the $\mathcal{P}^i$;

${L^i}_{boundary}$ is computed as the mean square error (MSE) between the predicted breast lesion boundary map (denoted as $\mathcal{D}^i$) and ground truth of the boundary map (denoted as $\mathcal{B}_G$):
\begin{equation}
\label{eq:loss-boundary}
{L^i}_{boundary} =\sum_{j=1}^{N_p}\{(\mathcal{D}^i)_j - (\mathcal{B}_G)_j\}^2 \ ,
\end{equation}
where ${N_p}$ is the number of pixel in $\mathcal{D}^i$;
$(\mathcal{D}^i)_j$ is the j-th pixel at $\mathcal{D}^i$;
and
$(\mathcal{B}_G)_j$ is the j-th pixel at $\mathcal{B}_G$.

Moreover, following ${L^i}_{seg}$, ${L^{f}}_{seg}$ also combines the dice coefficient loss and binary cross-entropy loss to compute the difference between the predicted segmentation map (denoted as $\mathcal{F}$) and $\mathcal{G}$ (see Eqn.~\ref{eq:loss-i-seg}):
\begin{equation}
\label{eq:final-seg-loss}
{L^{f}}_{seg} = 1- \dfrac{2\sum_{j=1}^{N_p} (\mathcal{F})_j \times (\mathcal{G})_j}{\sum_{j=1}^{N_p} (\mathcal{F})_j^2+\sum_{j=1}^{N_p} (\mathcal{G})_j^2} -\dfrac{1}{N_p}\sum_{j=1}^{N_p} (\mathcal{F})_j log (\mathcal{G})_j \ .
\end{equation}

\subsection{Implementation}
\subsubsection{Training Parameters}

To accelerate the training process, we initialize the parameters of the feature extract network using the pre-trained ResNext on ImageNet while other parameters are initialized by random noise.
The SGD algorithm is used to optimize the whole network with a momentum of 0.9, a weight decay of 0.0001, a mini-batch size of 4, and 100 epochs.
We set the initial learning rate as 0.001 and reduce it by multiplying 0.1 after finishing every 50 epochs.
Random rotation and horizontal flip operations are adopted for performing the data augmentation on the training set.
We implement the whole network using PyTorch library and train our network on a single NVIDIA TITIAN Xp GPU.

\subsubsection{Inference}
In the testing stage, we take the segmentation result predicted from the refined features of the dual guided non-local block as the output of our segmentation network, and then pass the result to the conditional random fields (CRF)~\citep{krahenbuhl2011efficient} for obtaining the final segmentation result. The network has 55M trainable parameters. The inference time was 0.039 seconds per image.


\begin{table*}[!t]
	\centering
	\caption{\revised{Quantitative results on our collected dataset and the number of parameters for all networks constructed in ablation study on our collected dataset.. The first row is Deeplabv3+ with ResNeXt as the feature extraction backbone. ``Guidance'' denotes guidance information. ``SNLB'' denotes the traditional spatial-wisely non-local block while ``SNLB+Guidance'' is our spatial-wise GGB (see Fig.~\ref{fig:sgnlb}). ``CNLB'' is the traditional channel-wisely non-local block while ``CNLB+Guidance'' is our channel-wise GGB (see Fig.~\ref{fig:cgnlb}).}}
	\label{tab:tab1}
	\begin{tabular}{cccc|c|ccccc}
	\toprule[1.5pt]
     SNLB      &     CNLB      &   Guidance   &     BD     & Parameters &         Jaccard \%         &       Dice \%        &       Accuracy \%       &        Recall \%        &      Precision \%       \\ \hline
              &              &              &              &
   53.4 M   &     73.4 $\pm$ 2.5      &81.5 $\pm$ 2.6           &     97.0 $\pm$ 0.3      &     78.9 $\pm$ 2.4      &     88.7 $\pm$ 3.0      \\
 $\checkmark$ &              &              &              &   53.5 M   &  77.6 $\pm$ 1.3         &      84.5 $\pm$ 1.2     &     97.2 $\pm$ 0.3      &     83.1 $\pm$ 1.6      &     90.7 $\pm$ 1.7      \\
              & $\checkmark$ &              &              &   53.5 M   & 77.5 $\pm$ 1.6          &  84.3 $\pm$ 1.3         &     97.2 $\pm$ 0.4      &     83.5 $\pm$ 1.7      &     90.8 $\pm$ 1.6      \\

 $\checkmark$ &              & $\checkmark$ &              &   53.9 M   &   78.1 $\pm$ 1.4        &     85.0 $\pm$ 1.3      &     97.3 $\pm$ 0.4      &     84.1 $\pm$ 0.2      &     91.0 $\pm$ 1.5      \\
              & $\checkmark$ & $\checkmark$ &              &   53.9 M   & 78.2 $\pm$ 1.4        &    85.2 $\pm$ 1.2         &     97.3 $\pm$ 0.4      &     84.5 $\pm$ 1.2      &     90.9 $\pm$ 1.5      \\
 $\checkmark$ & $\checkmark$ &              &              &   55.2 M   & 78.4 $\pm$ 1.6           &   85.4 $\pm$ 1.4       &     97.2 $\pm$ 0.4      &     84.9 $\pm$ 1.8      &     90.9 $\pm$ 1.7      \\
 $\checkmark$ & $\checkmark$ & $\checkmark$ &              &   55.4 M   &  78.8 $\pm$ 1.7         &       86.7 $\pm$ 1.2    &     97.3 $\pm$ 0.3      &     86.1 $\pm$ 1.7      &     91.2 $\pm$ 1.2      \\
 $\checkmark$ & $\checkmark$ & $\checkmark$ & $\checkmark$ &   55.4 M   & \textbf{79.1 $\pm$ 1.6} & \textbf{87.1 $\pm$ 1.4}  & \textbf{97.4 $\pm$ 0.3} & \textbf{86.6 $\pm$ 1.7} & \textbf{91.3 $\pm$ 1.0} \\
 \bottomrule[1.5pt]	\end{tabular}
\end{table*}

\begin{table*}[!t]
	\centering
	\caption{\revised{Quantitative results on our method and that with the BD module on the network output branch on our collected dataset.}}
	\label{tab:BD-output}
	\begin{tabular}{c|c|ccccc}
	\toprule[1.5pt]
     &         Dice \%         &       Jaccard \%        &       Accuracy \%       &        Recall \%        &      Precision \%       \\ \hline
  Our method& \textbf{87.1 $\pm$ 1.4} & \textbf{79.1 $\pm$ 1.6} & \textbf{97.4 $\pm$ 0.3} & \textbf{86.6 $\pm$ 1.7} & \textbf{91.3 $\pm$ 1.0} \\
  Ours-BD &     87.0 $\pm$ 1.3      &     \textbf{79.1 $\pm$ 1.5}      &     97.3 $\pm$ 0.4      &     86.4 $\pm$ 1.0      &     91.0 $\pm$ 1.5      \\
 \bottomrule[1.5pt]	\end{tabular}
\end{table*}

\begin{table*}[ht]
	\centering
	\caption{Quantitative comparisons of our network with and without an alternative deep supervision in BDBL.}
	\label{tab:alternative-DS}
	\resizebox{\textwidth}{!}{%
		\begin{tabular}{lccccccc}
			\toprule[1.5pt]
			& Jaccard $\%$                                       & Dice $\%$                                      & Accuracy $\%$                                    & Recall $\%$                                                                  & Precision $\%$   &HD & ABD                                  \\ \hline
			Ours-ADS       & 78.5 $\pm$ 1.7                        & 86.6 $\pm$ 1.5                          & 97.3 $\pm$ 0.3                         & 86.3 $\pm$ 1.2                             & 86.1 $\pm$ 1.5              &16.4 $\pm$ 2.3    &5.5 $\pm$0.8       \\
			GG-Net (our method)      & \textbf{79.1 $\pm$ 1.6}               & \textbf{87.1 $\pm$ 1.2}               &\textbf{97.4 $\pm$ 0.3 }                       & \textbf{86.6 $\pm$ 1.7}                                & \textbf{91.3 $\pm$ 1.0} &\textbf{16.2 $\pm$ 2.4}    &\textbf{5.3 $\pm$ 0.7}  \\   
			\bottomrule[1.5pt]
		\end{tabular}%
	
	}
\end{table*}

\begin{table*}[ht]
	\centering
	\caption{Comparing our method (GG-Net) with the state-of-the-art
		methods for beast lesion segmentation on our collected dataset. }
	\label{tab:tab2}
	\resizebox{\textwidth}{!}{%
		\begin{tabular}{lccccccc}
			\toprule[1.5pt]
			& Jaccard $\%$                                       & Dice $\%$                                      & Accuracy $\%$                                    & Recall $\%$                                                                  & Precision $\%$   &HD & ABD                                  \\ \hline
			U-Net~(\cite{ronneberger2015u})       & 69.3 $\pm$ 2.4                        & 78.0 $\pm$ 2.4                          & 96.5 $\pm$ 0.3                         & 76.9 $\pm$ 0.3                       & 85.6 $\pm$ 2.4           &25.1 $\pm$ 2.4     &8.1 $\pm$ 0.9                 \\
			U-Net++~(\cite{zhou2018unet++})     &  73.3 $\pm$ 2.1  & 82.1 $\pm$ 2.2  & 96.6 $\pm$ 0.4  & 81.1 $\pm$ 1.7   & 87.9 $\pm$ 2.6 &25.6 $\pm$ 4.0     &8.4 $\pm$ 1.2   \\
		
			TernausNet ~(\cite{iglovikov2018ternausnet}) & 73.7 $\pm$ 1.5  &82.2 $\pm$ 1.5  & 96.8 $\pm$ 0.3  & 82.1 $\pm$ 1.2  & 86.9 $\pm$ 0.2 &21.6 $\pm$ 2.6    &7.5 $\pm$ 0.9  5\\
			FPN~(\cite{lin2017feature})        & 77.2 $\pm$ 1.9 &  85.4 $\pm$ 1.7 &  97.1 $\pm$ 0.4 &  85.6 $\pm$ 1.8  & 89.1 $\pm$ 2.4  &18.1 $\pm$ 2.7    &6.1 $\pm$ 1.0 \\
			DeepLabv3+~(\cite{chen2018encoder}) & 73.4 $\pm$ 2.5                        & 81.5 $\pm$ 2.6                          & 97.0 $\pm$ 0.3                         & 78.9 $\pm$ 2.4                             & 88.7 $\pm$ 3.0              &22.3 $\pm$ 4.1    &7.9 $\pm$ 1.3            \\ \hline
			AG-Unet~(\cite{schlemper2019attention}) &74.1 $\pm$ 1.9 & 82.8 $\pm$ 1.9 & 96.6 $\pm$ 0.4 & 82.5 $\pm$ 2.3 & 87.3 $\pm$ 1.9 &24.1 $\pm$ 3.0    &7.8 $\pm$ 1.0 \\
			DAF~(\cite{wang2018deep}) &75.4 $\pm$ 1.9 & 83.6 $\pm$ 2.1 & 97.1 $\pm$ 0.4 &84.5 $\pm$ 2.3 & 86.6 $\pm$ 2.4 & 17.1 $\pm$ 2.3    &5.8 $\pm$ 0.9 \\
			\hline

			GG-Net (our method)      & \textbf{79.1 $\pm$ 1.6}               & \textbf{87.1 $\pm$ 1.2}               &\textbf{97.4 $\pm$ 0.3 }                       & \textbf{86.6 $\pm$ 1.7}                                & \textbf{91.3 $\pm$ 1.0} &\textbf{16.2 $\pm$ 2.4}    &\textbf{5.3 $\pm$ 0.7}  \\   
			\bottomrule[1.5pt]
		\end{tabular}%
	
	}
\end{table*}

\begin{table*}[ht]
	\centering
	\caption{Comparing our method (GG-Net) with the state-of-the-art
		methods for beast lesion segmentation on the BUSI dataset. Best results are marked with bold texts. }
	\label{tab:tabbusi}
	\resizebox{\textwidth}{!}{%
		\begin{tabular}{lccccccc}
			\toprule[1.5pt]
			& Jaccard $\%$                                       & Dice $\%$                                      & Accuracy $\%$                                    & Recall $\%$                                                                  & Precision $\%$      &HD &ABD                             \\ \hline
			U-Net~(\cite{ronneberger2015u})       & 64.1 $\pm$ 1.8                        & 73.3 $\pm$ 1.7                          & 95.9 $\pm$ 0.6                         & 70.4 $\pm$ 1.9                       & 83.3 $\pm$ 1.3          & 65.2 $\pm$ 4.7                       & 24.4 $\pm$ 2.3               \\
			U-Net++~(\cite{zhou2018unet++})     &  56.2 $\pm$ 1.7  & 66.0 $\pm$ 1.4  & 95.4 $\pm$ 0.4  & 62.8 $\pm$ 1.5   &78.2 $\pm$ 1.2 & 78.6$\pm$ 6.1                       & 31.8 $\pm$ 4.0  \\
			FPN~(\cite{lin2017feature})        & 72.2 $\pm$ 1.6 &   80.4 $\pm$ 1.6 &   95.9 $\pm$ 0.6 &  79.3 $\pm$ 1.3  &85.1 $\pm$ 1.5 & 47.6$\pm$ 5.8                       & 18.9 $\pm$ 2.6\\
			DeepLabv3+~(\cite{chen2018encoder}) & 68.2 $\pm$ 1.8                        & 77.2 $\pm$ 1.6                          & 96.3 $\pm$ 0.6                         & 74.4 $\pm$ 2.5                             & 84.8 $\pm$ 1.8       & 54.4$\pm$ 5.9                       & 22.4 $\pm$ 2.9                 \\
			SK-U-Net~(\cite{byra2020breast}) & -                        & 70.9                           & 95.6                         & -                             & -       & -                       & -                 \\
			DAF(~\cite{wang2018deep}) & 68.4 $\pm$ 3.1                        & 77.1 $\pm$ 3.1                          & 96.4 $\pm$ 0.6                         & 76.7 $\pm$ 3.8                             & 82.2 $\pm$ 3.1       & 46.9$\pm$ 8.1                       & 17.9 $\pm$ 4.7                   \\
			GG-Net (our method)       & \textbf{73.8 $\pm$ 1.1}               & \textbf{82.1 $\pm$ 1.1}               &\textbf{96.9 $\pm$ 0.5 }                       & \textbf{81.2 $\pm$ 1.6}                                & \textbf{86.5 $\pm$ 0.5} & \textbf{43.9$\pm$ 4.8}                       & \textbf{16.4 $\pm$ 2.2}  \\   
			\bottomrule[1.5pt]
		\end{tabular}%
	
	}
\end{table*}

\begin{table*}[ht]
	\centering
	\caption{\revised{Comparing our method (GG-Net) with the state-of-the-art
		methods for beast lesion segmentation on the BUSI dataset. (include normal data). Best results are marked with bold texts. }}
	\label{tab:tabbusi2}
	\resizebox{\textwidth}{!}{%
		\begin{tabular}{lccccccc}
			\toprule[1.5pt]
			& Jaccard $\%$                                       & Dice $\%$                                      & Accuracy $\%$                                    & Recall $\%$                                                                  & Precision $\%$  & HD      &ABD              \\ \hline
			U-Net~(\cite{ronneberger2015u})       & 51.2 $\pm$ 1.9                        & 58.8 $\pm$ 1.5                          & 96.3 $\pm$ 0.7                         & 56.1 $\pm$ 2.3                       & 68.1 $\pm$ 1.7           &67.1 $\pm$ 6.1     &24.7 $\pm$ 3.1     \\
			U-Net++~(\cite{zhou2018unet++})     &  44.5 $\pm$ 3.5  & 52.1 $\pm$ 3.7  & 95.9 $\pm$ 0.2  & 48.8 $\pm$ 4.7   &63.6 $\pm$ 2.4 &73.5 $\pm$ 5.0     &27.8 $\pm$ 2.0 \\
			FPN~(\cite{lin2017feature})        & 55.4 $\pm$ 2.1 &   63.0 $\pm$ 2.3 &   96.2 $\pm$ 0.4 &  62.1 $\pm$ 3.4  &68.3 $\pm$ 1.9  &56.8 $\pm$ 8.9     &21.2 $\pm$ 4.6 \\
			DeepLabv3+(~\cite{chen2018encoder}) & 54.3 $\pm$ 2.1                        & 62.1 $\pm$ 2.5                          & 96.4 $\pm$ 0.5                         & 59.2 $\pm$ 2.4                             & 63.6 $\pm$ 2.5       &55.5 $\pm$ 10.7     &21.3 $\pm$ 5.5        \\
			DAF(~\cite{wang2018deep}) & 55.8 $\pm$ 1.5                        & 62.8 $\pm$ 1.8                          & 96.6 $\pm$ 0.6                         & 62.8 $\pm$ 2.3                             & 66.5 $\pm$ 1.1       &52.8 $\pm$ 4.2     &20.3 $\pm$ 2.3                  \\
			GG-Net (our method)       & \textbf{56.6 $\pm$ 1.9}               & \textbf{64.1 $\pm$ 2.1}               &\textbf{96.6 $\pm$ 0.3 }                       & \textbf{63.3 $\pm$ 3.6}                                & \textbf{69.7 $\pm$ 0.4}  &\textbf{48.6 $\pm$ 7.2}     &\textbf{18.8 $\pm$ 3.3} \\   
			\bottomrule[1.5pt]
		\end{tabular}%
	
	}
\end{table*}

\section{Experiments}
We first introduce two datasets on breast ultrasound lesion segmentation and evaluation metrics, then conduct ablation studies to verify the major components of our network, as well as quantitatively and qualitatively compare our method against the state-of-the-art segmentation methods.

\subsection{Datasets}

We evaluate our segmentation network on two datasets including a public benchmark dataset (\textit{i.e.}, BUSI in ~\cite{walid2019datset}) and our collected dataset.
BUSI collected $780$ images from 600 female patients, with $437$ benign cases, $210$ benign masses, and $133$ normal cases. \revised{Note that the main purpose of breast lesion segmentation in the clinical usage is for the lesion assessment, tracking the lesion change, and identifying distribution and seriousness of lesions. As a result, clinicians usually screen the input ultrasound sample firstly to identify the lesion, and then conduct the breast lesion segmentation for clinical measurement.}
As a result, we remove the normal cases without breast lesion  masks to form the benchmark dataset, and adopt the three-fold cross-validation to test each segmentation method.

Our collected dataset has 632 clinical breast ultrasound images in total from 200 patients. The images are captured by different ultrasound imaging systems from Shenzhen Peoples Hospital and the Second Affiliated Hospital of Jinan University.
\revised{
We follow the widely-used annotation procedure of the medical image segmentation for annotating breast lesions. Firstly, three experienced radiologists are invited to annotate the breast lesion regions of each ultrasound image using a software interface developed via Matlab. Each radiologist used about two weeks to delineate all the breast lesion regions, and the segmentation ground truths of each image were then obtained based on inner- and intra-observer agreement of the three radiologists. Then, the final ground-truths were further refined by a senior radiologist with more than 10-year experience for quality control.}
To make the comparisons fair, we adopt the seven-fold cross-validation to test each segmentation method on this dataset.

\subsection{Evaluation Metrics}

We adopt seven commonly used metrics to quantitatively compare different methods on the breast lesion segmentation. They are Dice coefficient (denoted as Dice), Jaccard index, Recall, Precision, Accuracy, \revised{Hausdorff distance (denoted as HD) and average boundary distance (denoted as ABD)}.

\subsection{Ablation Analysis of our GG-Net}

\revised{In this section, we show the effectiveness of the principal components of our network, i.e., sptial-wise GGB, channel-wise GGB, and BD module in our network. And the ablation study experiments are mainly conducted on our collected dataset.}
The baseline (i.e., first row of Table~\ref{tab:tab1}) is constructed by removing both \revised{GGB} and the \revised{BD} module from our network. It is the original DeeplabV3+ network with ResNeXt as the backbone.

Table \ref{tab:tab1} shows the comparison results of our method with different components.
\revised{By comparing `SNLB', `CNLB' and baseline (first row of Table~\ref{tab:tab1}), we can see that learning the long-range dependencies has a superior performance in segmenting the breast lesion regions from the ultrasound images.
Then, `SNLB + Guidance ' (i.e., spatial-wise GGB) and `CNLB + Guidance' (i.e., channel-wise GGB) have better results than `SNLB' and `CNLB', showing that adding our MLIF guidance information into the spatial non-local and the channel non-local can help to capture the long-range position dependencies for the breast lesion segmentation.}
\revised{Moreover, the combination of the spatial-wise GGB and the channel-wise GGB has superior segmentation results over only using spatial-wise GGB or channel-wise GGB}, demonstrating that combining the spatial and channel information into learning guided non-local features can enhance the breast lesion segmentation performance.
Finally, our method with full components has the best segmentation accuracy, which means that the detected breast lesion boundaries in the \revised{BD} module of our network also contribute to the superior breast lesion segmentation performance.

\begin{figure*}[t!]
	\centering
	\includegraphics[width=0.98\textwidth]{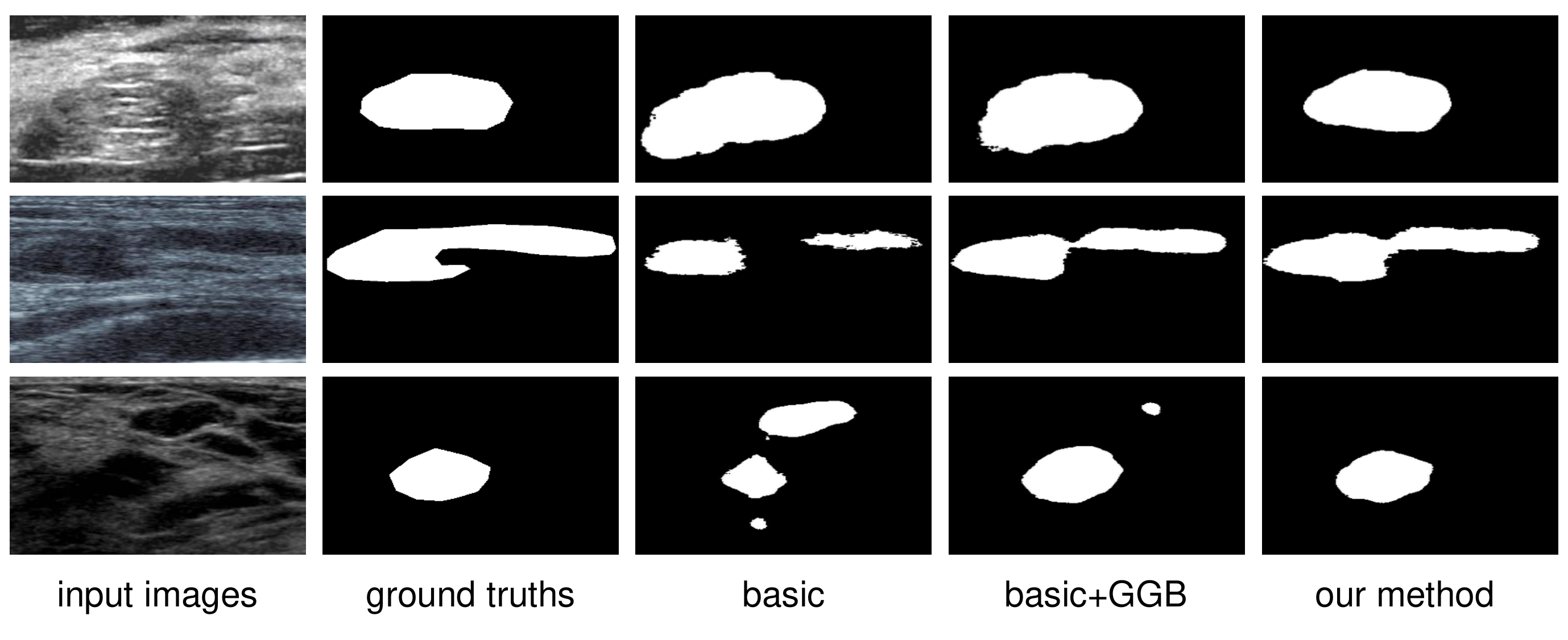}
	\caption{Visual results of ablation study. (a) Input images; (b) Ground truths; (c)-(e) are the segmentation produced by basic, \revised{``basic+GGB''}, and our method (i.e., ``basic+GGB+BD'') respectively.}
	\label{fig:5}
\end{figure*}

\revised{Fig.~\ref{fig:5} visually compares the segmentation results produced by the baseline, ``basic+GGB'' and our method.
From the visual results, we can easily find that ``basic+GGB'' has a higher segmentation accuracy than ``basic'', showing that the developed GGB can learn the long-range position dependencies to boost the breast lesion segmentation performance. Moreover, as shown in Fig.~\ref{fig:5} (e) and Fig.~\ref{fig:5} (d), our method (i.e., ``basic+GGB+BD'') can more accurately detect breast lesion regions than ``basic+GGB''.
It means that adding the BD module into our method can further improve the segmentation accuracy by generating refined boundaries.}

\revised{
\noindent
\textbf{BD on the network output branch.} \
Our network applies the BD module on different CNN layers; see Fig.~\ref{fig:2}. Here, we modify our network by applying the BD module on the output branch for detecting breast lesions and the modified network is denoted as ``Ours-BD''.
Table~\ref{tab:BD-output} lists different metric values of our method and ``Ours-BD'', showing that our method has only slightly better metric results than ``Ours-BD''. It means that adding the BD module on the network output branch reaches a similar segmentation accuracy as our network.
}

\revised{
\noindent
\textbf{Alternative deep supervision in BD modules.} \ Note that the BD module of our network imposes the deep supervision on two predictions, i.e., the breast lesion segmentation and the breast lesion boundary detection. To really verify the contribution of the BD module, we conduct an experiment by constructing a network (denoted as ‘Ours-ADS’) by using alternative deep supervision methods in the BD module, which means that we only impose the deep supervisions on the breast lesion segmentation and remove the supervisions on breast lesion boundary predictions in each BD module. Table~\ref{tab:alternative-DS} summarizes the quantitative results of our method and ‘Ours-ADS’ on our collected dataset. From the results, we can easily conclude that our method has achieved superior quantitative results than ‘Ours-ADS’ on all the seven evaluation metrics, demonstrating that utilizing an alternative deep supervision method (i.e., removing breast lesion boundary detection supervision) in the BD module reduces the breast lesion segmentation accuracy of our network.
}

\subsection{Comparison with the State-of-the-arts}

\noindent
\textbf{Compared methods.} \ We compare our network against several deep-learning-based segmentation methods, including \textit{context-based methods}: feature pyramid network (FPN) ~(\cite{lin2017feature}), U-Net~(\cite{ronneberger2015u}), U-Net++~(\cite{zhou2018unet++}), pre-trained TernausNet~(\cite{iglovikov2018ternausnet}),
\revised{SK-U-Net~(\cite{byra2020breast})},
DeeplabV3+~(\cite{chen2018encoder}); as well as \textit{attention-based methods}: AG-Unet~(\cite{schlemper2019attention}), and \revised{DAF~(\cite{wang2018deep}}).
To provide fair comparisons, we obtain the segmentation results of compared methods by downloading their public implementations and re-training their networks on our dataset.
Similarly, we also use the CRF~(\cite{krahenbuhl2011efficient}) to post-process the predicted segmentation maps of compared methods.

\noindent
\textbf{Quantitative comparisons.} \ \revised{Table~\ref{tab:tab2} reports the mean and standard deviation values of the seven metrics among our method and all the competitors on our collected dataset, while Table~\ref{tab:tabbusi} summarizes the mean and standard deviation scores of seven metrics on BUSI.
Compared to other segmentation methods,  our method has larger Jaccard, Dice, Accuracy, recall, and prediction scores, as well as smaller HD and ABD values.
It indicates that our GG-Net can more accurately identify breast lesions from ultrasound images than all the competitors.
}

\revised{
On the other hand, when further looking into the metric results in Table~\ref{tab:tab2} and Table~\ref{tab:tabbusi}, we can find that the segmentation performance on our collected dataset (see Table~\ref{tab:tab2}) is better than the results on the public BUSI dataset (seeTable~\ref{tab:tabbusi}) with respect to all seven evaluation metrics.
The reason behind is that the ultrasound image quality in our dataset is better than that in BUSI, thereby making the better segmentation performance.
}

\revised{
\vspace{2mm}
\noindent
\textbf{Utilizing BUSI's normal cases.} \  The general purpose of breast lesion segmentation in the clinical usage is mainly for the lesion assessment, tracking the lesion change, and identifying distribution and seriousness of lesions. As a result, people usually assume that the input ultrasound samples possess one or more lesions, and then conduct the breast lesion segmentation for clinical analysis. Here, we conduct another experiment by including the normal cases of BUSI into the training data and re-training all the compared methods and our network to obtain their new results. Table~\ref{tab:tabbusi} and Table~\ref{tab:tabbusi2} report the results of each method with and without the BUSI's normal cases. According to the results, we can easily find that the quantitative results of all the competitors and our network tend to be worse when considering normal cases in the network training. Among all the segmentation methods, our network still achieves the best performance of all seven metrics even though the normal cases are added into the training set and the testing set. }


\begin{figure*}[t!]
	\centering
	\includegraphics[width=\textwidth]{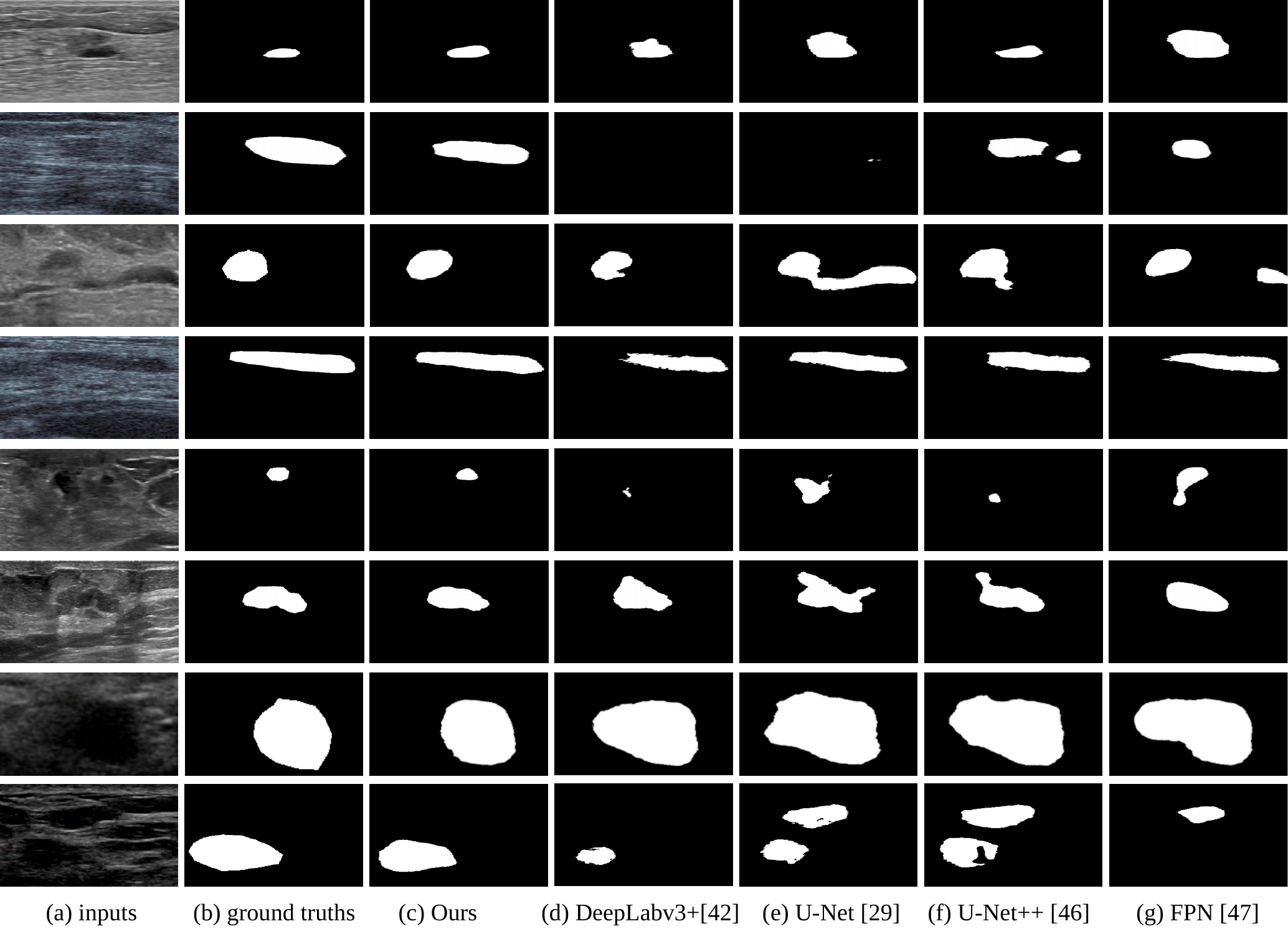}
	\caption{Visual comparison of the breast lesion segmentation maps produced by different methods. (a) input breast ultrasound images; (b) ground truths; (c)-(g) are segmentation results produced by our method, DeeplabV3+~\cite{chen2018encoder}, U-Net~\cite{ronneberger2015u}, U-Net++~\cite{zhou2018unet++}, and FPN~\cite{lin2017feature}. }
	\label{fig:6}
\end{figure*}

\noindent
\textbf{Visual comparisons.} We also visually compare the breast lesion segmentation results produced by our network and compared methods; see Fig. \ref{fig:6} for examples.
U-Net, U-Net++, FPN, and DeeplabV3+ tend to neglect breast lesion details or wrongly classify non-lesion regions as breast lesions into their predicted segmentation maps, while our method produces more accurate segmentation results on breast lesion regions.
Furthermore, our results are most consistent with ground truths (see Fig.~\ref{fig:6} (b)) among all segmentation results.
This proves the effectiveness of long-range dependencies and breast lesion boundaries in our method.

\section{Application}

Note that our network can be retrained for other ultrasound image segmentation tasks.
Hence, we further evaluate the effectiveness of our network by testing it on the ultrasound prostate segmentation task.
To conduct fair comparisons, we follow the same experimental setting of a recent prostate segmentation work, i.e., DAF~(\cite{wang2018deep}), to obtain the prostate segmentation results of our network.
We use the DAF's training set to train our network, test our method on the DAF's testing set,
and report the results of same four metric (i.e., Jaccard, Dice, Recall and Precision; see~(\cite{wang2018deep}) for their definitions) for comparisons.
Table~\ref{tab:tab4} summarizes the comparison results on four metrics between our method and state-of-the-art networks, including U-Net~(\cite{ronneberger2015u}), FCN~(\cite{lin2017feature}),  BCRNN~(\cite{yang2017fine}), and DAF~(\cite{wang2018deep}); see~(\cite{wang2018deep}) for details of these compared methods.
Apparently, our method outperforms all the competitors on almost all the four metrics, demonstrating that our method can also identify prostate regions better from ultrasound images.
It further verifies the effectiveness of the developed segmentation network in our work.

 \begin{table}[t]
 \caption{Metric results of different methods on ultrasound prostate segmentation.}
 \centering
 \label{tab:tab4}
 \resizebox{0.49\textwidth}{!}{
 \begin{tabular}{ccccc}
 \toprule[1.5pt]
           & Jaccard $\%$             & Dice $\%$              & Recall $\%$          & Precision $\%$                    \\ \hline
 FCN~(\cite{lin2017feature})       & 85.1         & 91.9          & 90.8          & 93.3                   \\
 BCRNN~(\cite{yang2017fine})     & 86.0          & 92.4          & 90.5          & 94.5          \\
 U-Net~(\cite{ronneberger2015u})     & 87.1          & 93.0          & 96.8          & 89.9           \\
 DAF~(\cite{wang2018deep})       & 91.0          & 95.3          & \textbf{97.0} & 93.7            \\
 GG-Net (ours) & \textbf{91.2} & \textbf{95.4} & 95.7          & \textbf{95.1}          \\
 \bottomrule[1.5pt]

 \end{tabular}
 }
 \end{table}

\begin{figure*}[t]
	\centering
    	\includegraphics[width=\textwidth]{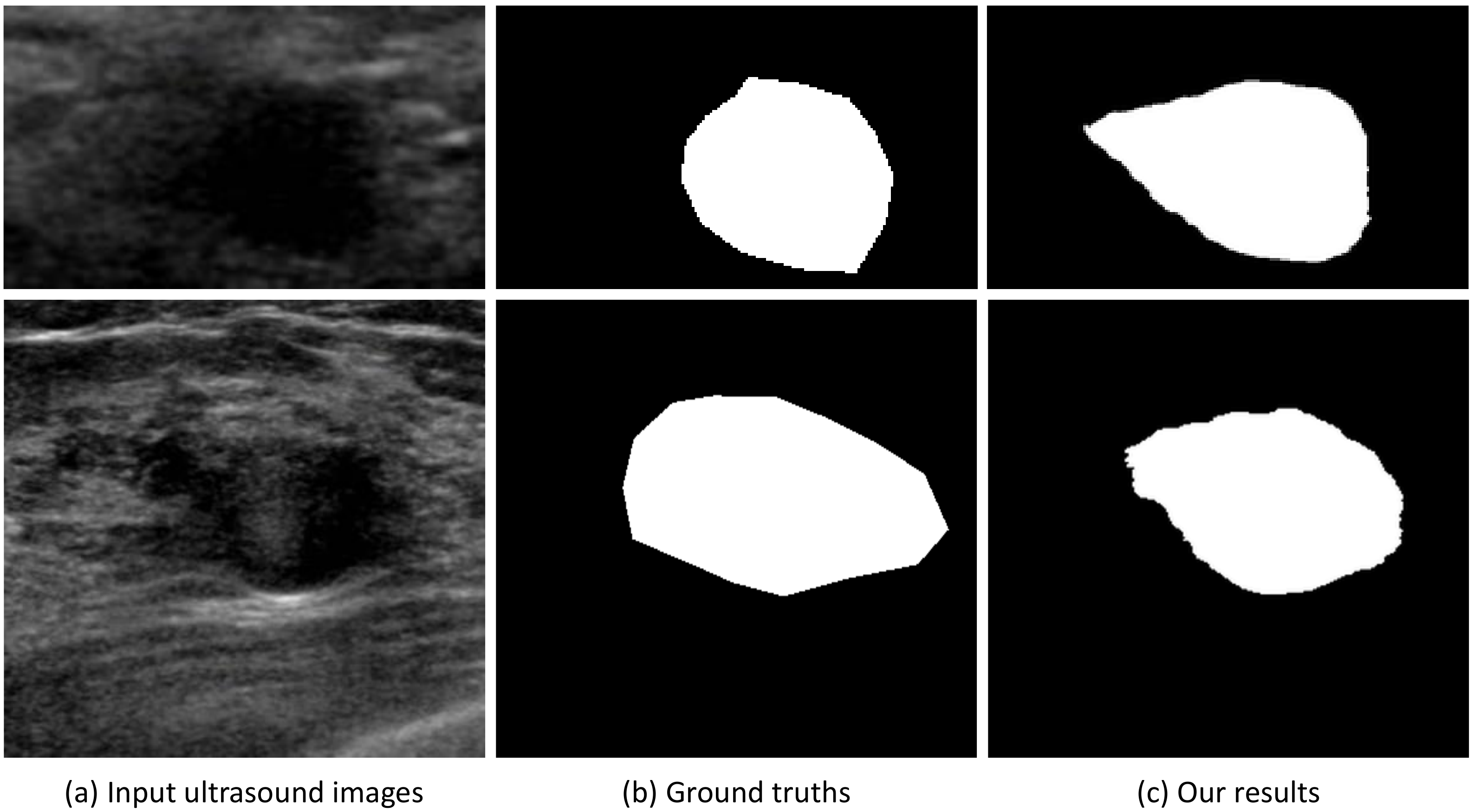}
	\caption{\revisedMR{Failure cases. (a) Input ultrasound images. (b) Ground truths of the breast lesion segmentation. (c) Segmentation results produced by our network.}}
	\label{fig:failure}
\end{figure*}

\begin{table*}[!t]
	\centering
	\caption{\revised{P-values between our method and other compared methods on different evaluation metrics.}}
	\label{tab:p-values}
	\resizebox{\linewidth}{!}{
	\begin{tabular}{c|c|c|c|c|c|c|c}
	\toprule[1.5pt]
     Metrics      &     U-Net vs Ours & U-Net++  vs Ours & TernausNet  vs Ours &FPN  vs Ours &AG-Net  vs Ours &DAF  vs Ours &DeepLabv3+  vs Ours      \\ \hline
Jaccard &1.62$\times$10$^{-7}$ &1.64$\times$10$^{-5}$
&9.41$\times$10$^{-5}$  &4.60$\times$10$^{-2}$  &3.45$\times$10$^{-4}$    &4.20$\times$10$^{-3}$ &3.00$\times$10$^{-6}$    \\
Dice &3.94$\times$10$^{-8}$ &2.81$\times$10$^{-5}$
&5.17$\times$10$^{-5}$  &4.00$\times$10$^{-2}$  &4.19$\times$10$^{-4}$    &1.10$\times$10$^{-3}$ &7.86$\times$10$^{-6}$    \\
Accuracy &1.74$\times$10$^{-3}$ &3.28$\times$10$^{-3}$
&3.48$\times$10$^{-2}$  &3.70$\times$10$^{-2}$  &1.33$\times$10$^{-3}$    &4.80$\times$10$^{-2}$ &3.57$\times$10$^{-2}$    \\
Recall &9.80$\times$10$^{-11}$ &2.31$\times$10$^{-4}$
&3.02$\times$10$^{-3}$  &3.50$\times$10$^{-2}$  &4.55$\times$10$^{-3}$    &1.20$\times$10$^{-3}$ &3.97$\times$10$^{-9}$    \\
Precision &6.70$\times$10$^{-3}$ &2.36$\times$10$^{-3}$
&3.24$\times$10$^{-4}$  &6.60$\times$10$^{-3}$  &2.06$\times$10$^{-3}$    &8.90$\times$10$^{-3}$ &2.70$\times$10$^{-3}$    \\
HD &1.64$\times$10$^{-6}$ &6.10$\times$10$^{-6}$
&7.35$\times$10$^{-4}$  &4.00$\times$10$^{-4}$  &2.89$\times$10$^{-6}$    &8.40$\times$10$^{-3}$ &4.00$\times$10$^{-4}$    \\
ABD &1.01$\times$10$^{-6}$ &1.73$\times$10$^{-7}$
&8.07$\times$10$^{-5}$  &2.00$\times$10$^{-3}$  &6.03$\times$10$^{-5}$    &5.50$\times$10$^{-3}$ &7.40$\times$10$^{-3}$    \\

\bottomrule[1.5pt]	\end{tabular}
  }
\end{table*}

\section{Discussions}
\revised{Breast cancer is the most frequently diagnosed cancer and the leading cause of cancer-related death among women worldwide.
The automatic breast lesion segmentation from ultrasound images assists the doctors in finding early signals of breast cancer, which is of great importance in clinical practice.
Traditional CNN-based methods~\citep{chen2018encoder,chen2017deeplab,chen2014semantic,ronneberger2015u,lin2017feature} conducted convolutional operations in local regions to learn deep discriminative features for medical image analysis and thus suffered from unsatisfactory segmentation accuracy due to the limited receptive fields of their local convolutions.

Recently, capturing non-local long-range pixel dependencies has achieved superior prediction performance in many medical imaging community~\citep{qi2019x,dou2018local} by devising non-local blocks.
However, these non-local blocks are only embedded into the deep CNN layers for network predictions. However, the deep layers of a segmentation network are responsible for capturing cues of the whole breast lesions and somehow lack parts of breast lesion regions due to the relatively larger receptive fields than shallow CNN layers. In this regard, we integrate all CNN layers to produce multi-level integrated features (MLIF) as a guidance information of the non-local blocks to complement more breast lesion boundary details (neglected by deep CNN layers).

This project presented a global guidance network (denoted as ``GG-Net'') with a spatial guidance block and a channel guidance block to leverage guidance information for improving long-range dependency feature learning in spatial and channel manners.  Moreover, a breast lesion boundary detection module is devised to learn boundary details for futher refining the breast lesion segmentation performance. Compared with state-of-the-art methods, our network achieves a significant (p-value $<$0.05, see Table~\ref{tab:p-values}) improvement on two datasets, which proves the effectiveness of the developed spatial and channel guidance block as well as boundary detection block. Moreover, compared with other segmentation networks, our method has better performance on relatively less obvious lesion segmentation. This is crucial in the clinical practice, especially for breast ultrasound, where most of the lesions have low contrast, shadows, and blurry boundaries.

Note that the elastography image encodes the density of the tissue in the screen. In our future work, we will leverage  elastography images for further boosting breast lesion segmentation results in ultrasound.
}

\revisedMR{
\vspace{2mm}
\noindent
\textbf{Failure cases.} \ Like other breast lesion segmentation methods, our network tends to fail in fully detecting breast lesion regions when the target breast lesion has a very large size and a complicated intensity distribution inside it, or unclear boundaries. Fig.~\ref{fig:failure} shows two examples, where our results in (c) wrongly identify non-lesion regions as lesion ones, or neglect a part of breast lesion regions of the input ultrasound image when comparing to the ground truths (see (b)).
}

\revised{
\vspace{2mm}
\noindent
\textbf{Statistical test.} \  To investigate the statistical significance of the proposed network over compared methods on different quantitative metrics, we conduct a statistical analysis of p-values and show the p-values of our network against compared methods in terms of different metrics in Table~\ref{tab:p-values}.
As shown in Table~\ref{tab:p-values}, we can find that the p-values of all the seven paired methods are almost smaller than 0.05 for all the seven metrics, demonstrating that our method can be
regarded as reaching a significant improvement over the other six compared methods on these evaluation metrics.
Note that the Accuracy p-values of our method over TernausNet, FPN, DAF, and DeepLabv3+ are 3.48$\times$10$^{-2}$, 3.70$\times$10$^{-2}$, 4.80$\times$10$^{-2}$, and 3.57$\times$10$^{-2}$, which are closer to 0.05. It indicates that our method has a similar Accuracy performance to TernausNet, FPN, DAF, and DeepLabv3+.
Generally, the superior metric performance of our method in Tables~\ref{tab:tab2}, ~\ref{tab:tabbusi}, and~\ref{tab:tabbusi2} shows that our network can better segment breast lesions from ultrasound than other compared segmentation methods.
}

\section{Conclusion}
\revised{This paper presents a global guidance network (GG-Net) equipped with a global guidance block and a breast lesion boundary detection module for breast lesion segmentation in ultrasound images.
The global guidance block aims to combine the multi-layer context information as guidance information to learn the long-term non-local features in spatial and channel manners. The breast lesion boundary detection predicts additional breast lesion boundary map to assist in improving the segmentation performance.}
%
%
We evaluate our network on a public dataset and our collected dataset of breast lesion segmentation in ultrasound images by comparing it against state-of-the-art methods, and the experimental results show that our network can more accurately segment the breast lesions than all the competitors.
We also show the application of our network on the ultrasound prostate segmentation task and our network also has a higher segmentation accuracy than state-of-the-art methods.

\bibliographystyle{model2-names.bst}\biboptions{authoryear}
\bibliography{reference}

\begin{thebibliography}{61}
\expandafter\ifx\csname natexlab\endcsname\relax\def\natexlab#1{#1}\fi
\providecommand{\url}[1]{\texttt{#1}}
\providecommand{\href}[2]{#2}
\providecommand{\path}[1]{#1}
\providecommand{\DOIprefix}{doi:}
\providecommand{\ArXivprefix}{arXiv:}
\providecommand{\URLprefix}{URL: }
\providecommand{\Pubmedprefix}{pmid:}
\providecommand{\doi}[1]{\href{http://dx.doi.org/#1}{\path{#1}}}
\providecommand{\Pubmed}[1]{\href{pmid:#1}{\path{#1}}}
\providecommand{\bibinfo}[2]{#2}
\ifx\xfnm\relax \def\xfnm[#1]{\unskip,\space#1}\fi
\bibitem[{Ahn et~al.(2017)Ahn, Heo, Jin and Kim}]{ahn2017novel}
\bibinfo{author}{Ahn, C.K.}, \bibinfo{author}{Heo, C.}, \bibinfo{author}{Jin,
  H.}, \bibinfo{author}{Kim, J.H.}, \bibinfo{year}{2017}.
\newblock \bibinfo{title}{A novel deep learning-based approach to high accuracy
  breast density estimation in digital mammography}, in:
  \bibinfo{booktitle}{Medical Imaging 2017: Computer-Aided Diagnosis},
  \bibinfo{organization}{International Society for Optics and Photonics}. p.
  \bibinfo{pages}{101342O}.
\bibitem[{Al-Dhabyani et~al.(2020)Al-Dhabyani, Gomaa, Khaled and
  FahmyaH}]{walid2019datset}
\bibinfo{author}{Al-Dhabyani, W.}, \bibinfo{author}{Gomaa, M.},
  \bibinfo{author}{Khaled, H.}, \bibinfo{author}{FahmyaH, A.},
  \bibinfo{year}{2020}.
\newblock \bibinfo{title}{Dataset of breast ultrasound images}.
\newblock \bibinfo{journal}{Data in Brief} \bibinfo{volume}{28}.
\bibitem[{{American Cancer Society}(2019)}]{American2019}
\bibinfo{author}{{American Cancer Society}}, \bibinfo{year}{2019}.
\newblock \bibinfo{title}{Cancer facts $\&$ figures 2019}.
\bibitem[{Ashton and Parker(1995)}]{ashton1995multiple}
\bibinfo{author}{Ashton, E.A.}, \bibinfo{author}{Parker, K.J.},
  \bibinfo{year}{1995}.
\newblock \bibinfo{title}{Multiple resolution bayesian segmentation of
  ultrasound images}.
\newblock \bibinfo{journal}{Ultrasonic imaging} \bibinfo{volume}{17},
  \bibinfo{pages}{291--304}.
\bibitem[{Boukerroui et~al.(1998)Boukerroui, Basset, Guerin and
  Baskurt}]{boukerroui1998multiresolution}
\bibinfo{author}{Boukerroui, D.}, \bibinfo{author}{Basset, O.},
  \bibinfo{author}{Guerin, N.}, \bibinfo{author}{Baskurt, A.},
  \bibinfo{year}{1998}.
\newblock \bibinfo{title}{Multiresolution texture based adaptive clustering
  algorithm for breast lesion segmentation}.
\newblock \bibinfo{journal}{European Journal of Ultrasound}
  \bibinfo{volume}{8}, \bibinfo{pages}{135--144}.
\bibitem[{Byra et~al.(2020)Byra, Jarosik, Szubert, Galperin, Ojeda-Fournier,
  Olson, O’Boyle, Comstock and Andre}]{byra2020breast}
\bibinfo{author}{Byra, M.}, \bibinfo{author}{Jarosik, P.},
  \bibinfo{author}{Szubert, A.}, \bibinfo{author}{Galperin, M.},
  \bibinfo{author}{Ojeda-Fournier, H.}, \bibinfo{author}{Olson, L.},
  \bibinfo{author}{O’Boyle, M.}, \bibinfo{author}{Comstock, C.},
  \bibinfo{author}{Andre, M.}, \bibinfo{year}{2020}.
\newblock \bibinfo{title}{Breast mass segmentation in ultrasound with selective
  kernel u-net convolutional neural network}.
\newblock \bibinfo{journal}{Biomedical Signal Processing and Control}
  \bibinfo{volume}{61}, \bibinfo{pages}{102027}.
\bibitem[{Canny(1986)}]{canny1986computational}
\bibinfo{author}{Canny, J.}, \bibinfo{year}{1986}.
\newblock \bibinfo{title}{A computational approach to edge detection}.
\newblock \bibinfo{journal}{IEEE Transactions on pattern analysis and machine
  intelligence} , \bibinfo{pages}{679--698}.
\bibitem[{Chen et~al.(2002)Chen, Lu and Huang}]{chen2002cell}
\bibinfo{author}{Chen, C.M.}, \bibinfo{author}{Lu, H.H.S.},
  \bibinfo{author}{Huang, Y.S.}, \bibinfo{year}{2002}.
\newblock \bibinfo{title}{Cell-based dual snake model: a new approach to
  extracting highly winding boundaries in the ultrasound images}.
\newblock \bibinfo{journal}{Ultrasound in medicine \& biology}
  \bibinfo{volume}{28}, \bibinfo{pages}{1061--1073}.
\bibitem[{Chen et~al.(2017a)Chen, Zhang, Xiao, Nie, Shao, Liu and
  Chua}]{chen2017sca}
\bibinfo{author}{Chen, L.}, \bibinfo{author}{Zhang, H.}, \bibinfo{author}{Xiao,
  J.}, \bibinfo{author}{Nie, L.}, \bibinfo{author}{Shao, J.},
  \bibinfo{author}{Liu, W.}, \bibinfo{author}{Chua, T.S.},
  \bibinfo{year}{2017}a.
\newblock \bibinfo{title}{{SCA-CNN}: Spatial and channel-wise attention in
  convolutional networks for image captioning}, in:
  \bibinfo{booktitle}{Proceedings of the IEEE conference on computer vision and
  pattern recognition}, pp. \bibinfo{pages}{5659--5667}.
\bibitem[{Chen et~al.(2014)Chen, Papandreou, Kokkinos, Murphy and
  Yuille}]{chen2014semantic}
\bibinfo{author}{Chen, L.C.}, \bibinfo{author}{Papandreou, G.},
  \bibinfo{author}{Kokkinos, I.}, \bibinfo{author}{Murphy, K.},
  \bibinfo{author}{Yuille, A.L.}, \bibinfo{year}{2014}.
\newblock \bibinfo{title}{Semantic image segmentation with deep convolutional
  nets and fully connected crfs}.
\newblock \bibinfo{journal}{arXiv preprint arXiv:1412.7062} .
\bibitem[{Chen et~al.(2017b)Chen, Papandreou, Kokkinos, Murphy and
  Yuille}]{chen2017deeplab}
\bibinfo{author}{Chen, L.C.}, \bibinfo{author}{Papandreou, G.},
  \bibinfo{author}{Kokkinos, I.}, \bibinfo{author}{Murphy, K.},
  \bibinfo{author}{Yuille, A.L.}, \bibinfo{year}{2017}b.
\newblock \bibinfo{title}{Deeplab: Semantic image segmentation with deep
  convolutional nets, atrous convolution, and fully connected crfs}.
\newblock \bibinfo{journal}{IEEE transactions on pattern analysis and machine
  intelligence} \bibinfo{volume}{40}, \bibinfo{pages}{834--848}.
\bibitem[{Chen et~al.(2018)Chen, Zhu, Papandreou, Schroff and
  Adam}]{chen2018encoder}
\bibinfo{author}{Chen, L.C.}, \bibinfo{author}{Zhu, Y.},
  \bibinfo{author}{Papandreou, G.}, \bibinfo{author}{Schroff, F.},
  \bibinfo{author}{Adam, H.}, \bibinfo{year}{2018}.
\newblock \bibinfo{title}{Encoder-decoder with atrous separable convolution for
  semantic image segmentation}, in: \bibinfo{booktitle}{Proceedings of the
  European conference on computer vision (ECCV)}, pp.
  \bibinfo{pages}{801--818}.
\bibitem[{Dhungel et~al.(2017)Dhungel, Carneiro and Bradley}]{dhungel2017deep}
\bibinfo{author}{Dhungel, N.}, \bibinfo{author}{Carneiro, G.},
  \bibinfo{author}{Bradley, A.P.}, \bibinfo{year}{2017}.
\newblock \bibinfo{title}{A deep learning approach for the analysis of masses
  in mammograms with minimal user intervention}.
\newblock \bibinfo{journal}{Medical Image Analysis} \bibinfo{volume}{37},
  \bibinfo{pages}{114--128}.
\bibitem[{Dou et~al.(2016)Dou, Chen, Jin, Yu, Qin and Heng}]{dou20163d}
\bibinfo{author}{Dou, Q.}, \bibinfo{author}{Chen, H.}, \bibinfo{author}{Jin,
  Y.}, \bibinfo{author}{Yu, L.}, \bibinfo{author}{Qin, J.},
  \bibinfo{author}{Heng, P.A.}, \bibinfo{year}{2016}.
\newblock \bibinfo{title}{3{D} deeply supervised network for automatic liver
  segmentation from ct volumes}, in: \bibinfo{booktitle}{International
  Conference on Medical Image Computing and Computer-Assisted Intervention
  (MICCAI)}, \bibinfo{organization}{Springer}. pp. \bibinfo{pages}{149--157}.
\bibitem[{Dou et~al.(2018)Dou, Zhang, Zheng and Zhou}]{dou2018local}
\bibinfo{author}{Dou, T.}, \bibinfo{author}{Zhang, L.}, \bibinfo{author}{Zheng,
  H.}, \bibinfo{author}{Zhou, W.}, \bibinfo{year}{2018}.
\newblock \bibinfo{title}{Local and non-local deep feature fusion for
  malignancy characterization of hepatocellular carcinoma}, in:
  \bibinfo{booktitle}{International Conference on Medical Image Computing and
  Computer-Assisted Intervention (MICCAI)}, \bibinfo{organization}{Springer}.
  pp. \bibinfo{pages}{472--479}.
\bibitem[{Feng et~al.(2019)Feng, Lu and Ding}]{feng2019attentive}
\bibinfo{author}{Feng, M.}, \bibinfo{author}{Lu, H.}, \bibinfo{author}{Ding,
  E.}, \bibinfo{year}{2019}.
\newblock \bibinfo{title}{Attentive feedback network for boundary-aware salient
  object detection}, in: \bibinfo{booktitle}{Proceedings of the IEEE Conference
  on Computer Vision and Pattern Recognition}, pp. \bibinfo{pages}{1623--1632}.
\bibitem[{Heinrich et~al.(2019)Heinrich, Oktay and
  Bouteldja}]{heinrich2019obelisk}
\bibinfo{author}{Heinrich, M.P.}, \bibinfo{author}{Oktay, O.},
  \bibinfo{author}{Bouteldja, N.}, \bibinfo{year}{2019}.
\newblock \bibinfo{title}{Obelisk-{N}et: Fewer layers to solve 3{D} multi-organ
  segmentation with sparse deformable convolutions}.
\newblock \bibinfo{journal}{Medical image analysis} \bibinfo{volume}{54},
  \bibinfo{pages}{1--9}.
\bibitem[{Hou et~al.(2019)Hou, Ma, Chang, Gu, Shan and
  Chen}]{hou2019interaction}
\bibinfo{author}{Hou, R.}, \bibinfo{author}{Ma, B.}, \bibinfo{author}{Chang,
  H.}, \bibinfo{author}{Gu, X.}, \bibinfo{author}{Shan, S.},
  \bibinfo{author}{Chen, X.}, \bibinfo{year}{2019}.
\newblock \bibinfo{title}{Interaction-and-aggregation network for person
  re-identification}, in: \bibinfo{booktitle}{Proceedings of the IEEE
  conference on computer vision and pattern recognition}, pp.
  \bibinfo{pages}{9317--9326}.
\bibitem[{Hu et~al.(2018)Hu, Shen and Sun}]{hu2018squeeze}
\bibinfo{author}{Hu, J.}, \bibinfo{author}{Shen, L.}, \bibinfo{author}{Sun,
  G.}, \bibinfo{year}{2018}.
\newblock \bibinfo{title}{Squeeze-and-excitation networks}, in:
  \bibinfo{booktitle}{Proceedings of the IEEE conference on computer vision and
  pattern recognition}, pp. \bibinfo{pages}{7132--7141}.
\bibitem[{Hu et~al.(2019)Hu, Guo, Wang, Yu, Li, Zhou and
  Chang}]{hu2019automatic}
\bibinfo{author}{Hu, Y.}, \bibinfo{author}{Guo, Y.}, \bibinfo{author}{Wang,
  Y.}, \bibinfo{author}{Yu, J.}, \bibinfo{author}{Li, J.},
  \bibinfo{author}{Zhou, S.}, \bibinfo{author}{Chang, C.},
  \bibinfo{year}{2019}.
\newblock \bibinfo{title}{Automatic tumor segmentation in breast ultrasound
  images using a dilated fully convolutional network combined with an active
  contour model}.
\newblock \bibinfo{journal}{Medical physics} \bibinfo{volume}{46},
  \bibinfo{pages}{215--228}.
\bibitem[{Huang et~al.(2008)Huang, Chen and Moon}]{huang2008neural}
\bibinfo{author}{Huang, S.F.}, \bibinfo{author}{Chen, Y.C.},
  \bibinfo{author}{Moon, W.K.}, \bibinfo{year}{2008}.
\newblock \bibinfo{title}{Neural network analysis applied to tumor segmentation
  on 3{D} breast ultrasound images}, in: \bibinfo{booktitle}{IEEE International
  Symposium on Biomedical Imaging: From Nano to Macro}, pp.
  \bibinfo{pages}{1303--1306}.
\bibitem[{Iglovikov and Shvets(2018)}]{iglovikov2018ternausnet}
\bibinfo{author}{Iglovikov, V.}, \bibinfo{author}{Shvets, A.},
  \bibinfo{year}{2018}.
\newblock \bibinfo{title}{Ternausnet: {U-Net} with vgg11 encoder pre-trained on
  imagenet for image segmentation}.
\newblock \bibinfo{journal}{arXiv preprint arXiv:1801.05746} .
\bibitem[{Joutard et~al.(2019)Joutard, Dorent, Isaac, Ourselin, Vercauteren and
  Modat}]{joutard2019permutohedral}
\bibinfo{author}{Joutard, S.}, \bibinfo{author}{Dorent, R.},
  \bibinfo{author}{Isaac, A.}, \bibinfo{author}{Ourselin, S.},
  \bibinfo{author}{Vercauteren, T.}, \bibinfo{author}{Modat, M.},
  \bibinfo{year}{2019}.
\newblock \bibinfo{title}{Permutohedral attention module for efficient
  non-local neural networks}, in: \bibinfo{booktitle}{International Conference
  on Medical Image Computing and Computer-Assisted Intervention},
  \bibinfo{organization}{Springer}. pp. \bibinfo{pages}{393--401}.
\bibitem[{Kirberger(1995)}]{kirberger1995imaging}
\bibinfo{author}{Kirberger, R.M.}, \bibinfo{year}{1995}.
\newblock \bibinfo{title}{Imaging artifacts in diagnostic ultrasound - a
  review}.
\newblock \bibinfo{journal}{Veterinary Radiology \& Ultrasound}
  \bibinfo{volume}{36}, \bibinfo{pages}{297--306}.
\bibitem[{Kr{\"a}henb{\"u}hl and Koltun(2011)}]{krahenbuhl2011efficient}
\bibinfo{author}{Kr{\"a}henb{\"u}hl, P.}, \bibinfo{author}{Koltun, V.},
  \bibinfo{year}{2011}.
\newblock \bibinfo{title}{Efficient inference in fully connected crfs with
  gaussian edge potentials}, in: \bibinfo{booktitle}{Advances in neural
  information processing systems}, pp. \bibinfo{pages}{109--117}.
\bibitem[{Kwak et~al.(2005)Kwak, Kim and Kim}]{kwak2005rd}
\bibinfo{author}{Kwak, J.I.}, \bibinfo{author}{Kim, S.H.},
  \bibinfo{author}{Kim, N.C.}, \bibinfo{year}{2005}.
\newblock \bibinfo{title}{R{D}-based seeded region growing for extraction of
  breast tumor in an ultrasound volume}, in: \bibinfo{booktitle}{International
  Conference on Computational and Information Science},
  \bibinfo{organization}{Springer}. pp. \bibinfo{pages}{799--808}.
\bibitem[{Lei et~al.(2018)Lei, Huang, Li, Bian, Li, Chou and
  Cheng}]{lei2018segmentation}
\bibinfo{author}{Lei, B.}, \bibinfo{author}{Huang, S.}, \bibinfo{author}{Li,
  R.}, \bibinfo{author}{Bian, C.}, \bibinfo{author}{Li, H.},
  \bibinfo{author}{Chou, Y.H.}, \bibinfo{author}{Cheng, J.Z.},
  \bibinfo{year}{2018}.
\newblock \bibinfo{title}{Segmentation of breast anatomy for automated whole
  breast ultrasound images with boundary regularized convolutional
  encoder--decoder network}.
\newblock \bibinfo{journal}{Neurocomputing} \bibinfo{volume}{321},
  \bibinfo{pages}{178--186}.
\bibitem[{Li et~al.(2018)Li, He, Zhou, Yu, Ni, Chen, Wang and
  Lei}]{li2018dense}
\bibinfo{author}{Li, H.}, \bibinfo{author}{He, X.}, \bibinfo{author}{Zhou, F.},
  \bibinfo{author}{Yu, Z.}, \bibinfo{author}{Ni, D.}, \bibinfo{author}{Chen,
  S.}, \bibinfo{author}{Wang, T.}, \bibinfo{author}{Lei, B.},
  \bibinfo{year}{2018}.
\newblock \bibinfo{title}{Dense deconvolutional network for skin lesion
  segmentation}.
\newblock \bibinfo{journal}{IEEE Journal of Biomedical and Health Informatics}
  \bibinfo{volume}{23}, \bibinfo{pages}{527--537}.
\bibitem[{Lin et~al.(2017)Lin, Doll{\'a}r, Girshick, He, Hariharan and
  Belongie}]{lin2017feature}
\bibinfo{author}{Lin, T.Y.}, \bibinfo{author}{Doll{\'a}r, P.},
  \bibinfo{author}{Girshick, R.}, \bibinfo{author}{He, K.},
  \bibinfo{author}{Hariharan, B.}, \bibinfo{author}{Belongie, S.},
  \bibinfo{year}{2017}.
\newblock \bibinfo{title}{Feature pyramid networks for object detection}, in:
  \bibinfo{booktitle}{Proceedings of the IEEE conference on computer vision and
  pattern recognition}, pp. \bibinfo{pages}{2117--2125}.
\bibitem[{Liu et~al.(2010)Liu, Cheng, Huang, Tian, Tang and Liu}]{liu2010fully}
\bibinfo{author}{Liu, B.}, \bibinfo{author}{Cheng, H.D.},
  \bibinfo{author}{Huang, J.}, \bibinfo{author}{Tian, J.},
  \bibinfo{author}{Tang, X.}, \bibinfo{author}{Liu, J.}, \bibinfo{year}{2010}.
\newblock \bibinfo{title}{Fully automatic and segmentation-robust
  classification of breast tumors based on local texture analysis of ultrasound
  images}.
\newblock \bibinfo{journal}{Pattern Recognition} \bibinfo{volume}{43},
  \bibinfo{pages}{280--298}.
\bibitem[{Lo et~al.(2014)Lo, Shen, Huang and Chang}]{lo2014computer}
\bibinfo{author}{Lo, C.}, \bibinfo{author}{Shen, Y.W.}, \bibinfo{author}{Huang,
  C.S.}, \bibinfo{author}{Chang, R.F.}, \bibinfo{year}{2014}.
\newblock \bibinfo{title}{Computer-aided multiview tumor detection for
  automated whole breast ultrasound}.
\newblock \bibinfo{journal}{Ultrasonic imaging} \bibinfo{volume}{36},
  \bibinfo{pages}{3--17}.
\bibitem[{Madabhushi and Metaxas(2002)}]{madabhushi2002automatic}
\bibinfo{author}{Madabhushi, A.}, \bibinfo{author}{Metaxas, D.},
  \bibinfo{year}{2002}.
\newblock \bibinfo{title}{Automatic boundary extraction of ultrasonic breast
  lesions}, in: \bibinfo{booktitle}{Proceedings IEEE International Symposium on
  Biomedical Imaging}, \bibinfo{organization}{IEEE}. pp.
  \bibinfo{pages}{601--604}.
\bibitem[{Madabhushi and Metaxas(2003)}]{madabhushi2003combining}
\bibinfo{author}{Madabhushi, A.}, \bibinfo{author}{Metaxas, D.N.},
  \bibinfo{year}{2003}.
\newblock \bibinfo{title}{Combining low-, high-level and empirical domain
  knowledge for automated segmentation of ultrasonic breast lesions}.
\newblock \bibinfo{journal}{IEEE Transactions on Medical Imaging}
  \bibinfo{volume}{22}, \bibinfo{pages}{155--169}.
\bibitem[{Mishra et~al.(2018)Mishra, Chaudhury, Sarkar and
  Soin}]{mishra2018ultrasound}
\bibinfo{author}{Mishra, D.}, \bibinfo{author}{Chaudhury, S.},
  \bibinfo{author}{Sarkar, M.}, \bibinfo{author}{Soin, A.S.},
  \bibinfo{year}{2018}.
\newblock \bibinfo{title}{Ultrasound image segmentation: A deeply supervised
  network with attention to boundaries}.
\newblock \bibinfo{journal}{IEEE Transactions on Biomedical Engineering}
  \bibinfo{volume}{66}, \bibinfo{pages}{1637--1648}.
\bibitem[{Moon et~al.(2014)Moon, Lo, Chen, Shen, Chang, Huang, Chen, Hsu and
  Chang}]{moon2014tumor}
\bibinfo{author}{Moon, W.K.}, \bibinfo{author}{Lo, C.M.},
  \bibinfo{author}{Chen, R.T.}, \bibinfo{author}{Shen, Y.W.},
  \bibinfo{author}{Chang, J.M.}, \bibinfo{author}{Huang, C.S.},
  \bibinfo{author}{Chen, J.H.}, \bibinfo{author}{Hsu, W.W.},
  \bibinfo{author}{Chang, R.F.}, \bibinfo{year}{2014}.
\newblock \bibinfo{title}{Tumor detection in automated breast ultrasound images
  using quantitative tissue clustering}.
\newblock \bibinfo{journal}{Medical physics} \bibinfo{volume}{41},
  \bibinfo{pages}{042901}.
\bibitem[{Mordang et~al.(2016a)Mordang, Gubern-M{\'e}rida, den Heeten and
  Karssemeijer}]{mordang2016reducing}
\bibinfo{author}{Mordang, J.J.}, \bibinfo{author}{Gubern-M{\'e}rida, A.},
  \bibinfo{author}{den Heeten, G.}, \bibinfo{author}{Karssemeijer, N.},
  \bibinfo{year}{2016}a.
\newblock \bibinfo{title}{Reducing false positives of microcalcification
  detection systems by removal of breast arterial calcifications}.
\newblock \bibinfo{journal}{Medical physics} \bibinfo{volume}{43},
  \bibinfo{pages}{1676--1687}.
\bibitem[{Mordang et~al.(2016b)Mordang, Janssen, Bria, Kooi, Gubern-M{\'e}rida
  and Karssemeijer}]{mordang2016automatic}
\bibinfo{author}{Mordang, J.J.}, \bibinfo{author}{Janssen, T.},
  \bibinfo{author}{Bria, A.}, \bibinfo{author}{Kooi, T.},
  \bibinfo{author}{Gubern-M{\'e}rida, A.}, \bibinfo{author}{Karssemeijer, N.},
  \bibinfo{year}{2016}b.
\newblock \bibinfo{title}{Automatic microcalcification detection in
  multi-vendor mammography using convolutional neural networks}, in:
  \bibinfo{booktitle}{International Workshop on Breast Imaging},
  \bibinfo{organization}{Springer}. pp. \bibinfo{pages}{35--42}.
\bibitem[{Othman and Tizhoosh(2011)}]{othman2011segmentation}
\bibinfo{author}{Othman, A.A.}, \bibinfo{author}{Tizhoosh, H.R.},
  \bibinfo{year}{2011}.
\newblock \bibinfo{title}{Segmentation of breast ultrasound images using neural
  networks}, in: \bibinfo{booktitle}{Engineering Applications of Neural
  Networks}. \bibinfo{publisher}{Springer}, pp. \bibinfo{pages}{260--269}.
\bibitem[{Peng et~al.(2017)Peng, Zhang, Yu, Luo and Sun}]{peng2017large}
\bibinfo{author}{Peng, C.}, \bibinfo{author}{Zhang, X.}, \bibinfo{author}{Yu,
  G.}, \bibinfo{author}{Luo, G.}, \bibinfo{author}{Sun, J.},
  \bibinfo{year}{2017}.
\newblock \bibinfo{title}{Large kernel matters--improve semantic segmentation
  by global convolutional network}, in: \bibinfo{booktitle}{Proceedings of the
  IEEE conference on computer vision and pattern recognition}, pp.
  \bibinfo{pages}{4353--4361}.
\bibitem[{Qi et~al.(2019)Qi, Yang, Li, Liu, Wang, Liu and Wang}]{qi2019x}
\bibinfo{author}{Qi, K.}, \bibinfo{author}{Yang, H.}, \bibinfo{author}{Li, C.},
  \bibinfo{author}{Liu, Z.}, \bibinfo{author}{Wang, M.}, \bibinfo{author}{Liu,
  Q.}, \bibinfo{author}{Wang, S.}, \bibinfo{year}{2019}.
\newblock \bibinfo{title}{X-{N}et: Brain stroke lesion segmentation based on
  depthwise separable convolution and long-range dependencies}.
\newblock \bibinfo{journal}{arXiv preprint arXiv:1907.07000} .
\bibitem[{Ronneberger et~al.(2015)Ronneberger, Fischer and
  Brox}]{ronneberger2015u}
\bibinfo{author}{Ronneberger, O.}, \bibinfo{author}{Fischer, P.},
  \bibinfo{author}{Brox, T.}, \bibinfo{year}{2015}.
\newblock \bibinfo{title}{{U-Net}: Convolutional networks for biomedical image
  segmentation}, in: \bibinfo{booktitle}{International Conference on Medical
  Image Computing and Computer-Assisted Intervention (MICCAI)},
  \bibinfo{organization}{Springer}. pp. \bibinfo{pages}{234--241}.
\bibitem[{Roy et~al.(2018)Roy, Navab and Wachinger}]{roy2018recalibrating}
\bibinfo{author}{Roy, A.G.}, \bibinfo{author}{Navab, N.},
  \bibinfo{author}{Wachinger, C.}, \bibinfo{year}{2018}.
\newblock \bibinfo{title}{Recalibrating fully convolutional networks with
  spatial and channel “squeeze and excitation” blocks}.
\newblock \bibinfo{journal}{IEEE transactions on medical imaging}
  \bibinfo{volume}{38}, \bibinfo{pages}{540--549}.
\bibitem[{Schlemper et~al.(2019)Schlemper, Oktay, Schaap, Heinrich, Kainz,
  Glocker and Rueckert}]{schlemper2019attention}
\bibinfo{author}{Schlemper, J.}, \bibinfo{author}{Oktay, O.},
  \bibinfo{author}{Schaap, M.}, \bibinfo{author}{Heinrich, M.},
  \bibinfo{author}{Kainz, B.}, \bibinfo{author}{Glocker, B.},
  \bibinfo{author}{Rueckert, D.}, \bibinfo{year}{2019}.
\newblock \bibinfo{title}{Attention gated networks: Learning to leverage
  salient regions in medical images}.
\newblock \bibinfo{journal}{Medical image analysis} \bibinfo{volume}{53},
  \bibinfo{pages}{197--207}.
\bibitem[{Shan et~al.(2012)Shan, Cheng and Wang}]{shan2012completely}
\bibinfo{author}{Shan, J.}, \bibinfo{author}{Cheng, H.}, \bibinfo{author}{Wang,
  Y.}, \bibinfo{year}{2012}.
\newblock \bibinfo{title}{Completely automated segmentation approach for breast
  ultrasound images using multiple-domain features}.
\newblock \bibinfo{journal}{Ultrasound in medicine \& biology}
  \bibinfo{volume}{38}, \bibinfo{pages}{262--275}.
\bibitem[{Shan et~al.(2008)Shan, Cheng and Wang}]{shan2008novel}
\bibinfo{author}{Shan, J.}, \bibinfo{author}{Cheng, H.D.},
  \bibinfo{author}{Wang, Y.}, \bibinfo{year}{2008}.
\newblock \bibinfo{title}{A novel automatic seed point selection algorithm for
  breast ultrasound images}, in: \bibinfo{booktitle}{2008 19th International
  Conference on Pattern Recognition}, \bibinfo{organization}{IEEE}. pp.
  \bibinfo{pages}{1--4}.
\bibitem[{Vaswani et~al.(2017)Vaswani, Shazeer, Parmar, Uszkoreit, Jones,
  Gomez, Kaiser and Polosukhin}]{vaswani2017attention}
\bibinfo{author}{Vaswani, A.}, \bibinfo{author}{Shazeer, N.},
  \bibinfo{author}{Parmar, N.}, \bibinfo{author}{Uszkoreit, J.},
  \bibinfo{author}{Jones, L.}, \bibinfo{author}{Gomez, A.N.},
  \bibinfo{author}{Kaiser, {\L}.}, \bibinfo{author}{Polosukhin, I.},
  \bibinfo{year}{2017}.
\newblock \bibinfo{title}{Attention is all you need}, in:
  \bibinfo{booktitle}{Advances in neural information processing systems}, pp.
  \bibinfo{pages}{5998--6008}.
\bibitem[{Wang et~al.(2018a)Wang, Girshick, Gupta and He}]{wang2018non}
\bibinfo{author}{Wang, X.}, \bibinfo{author}{Girshick, R.},
  \bibinfo{author}{Gupta, A.}, \bibinfo{author}{He, K.}, \bibinfo{year}{2018}a.
\newblock \bibinfo{title}{Non-local neural networks}, in:
  \bibinfo{booktitle}{Proceedings of the IEEE conference on computer vision and
  pattern recognition}, pp. \bibinfo{pages}{7794--7803}.
\bibitem[{Wang et~al.(2018b)Wang, Deng, Hu, Zhu, Yang, Xu, Heng and
  Ni}]{wang2018deep}
\bibinfo{author}{Wang, Y.}, \bibinfo{author}{Deng, Z.}, \bibinfo{author}{Hu,
  X.}, \bibinfo{author}{Zhu, L.}, \bibinfo{author}{Yang, X.},
  \bibinfo{author}{Xu, X.}, \bibinfo{author}{Heng, P.A.}, \bibinfo{author}{Ni,
  D.}, \bibinfo{year}{2018}b.
\newblock \bibinfo{title}{Deep attentional features for prostate segmentation
  in ultrasound}, in: \bibinfo{booktitle}{International Conference on Medical
  Image Computing and Computer-Assisted Intervention (MICCAI)},
  \bibinfo{organization}{Springer}. pp. \bibinfo{pages}{523--530}.
\bibitem[{Xian et~al.(2015)Xian, Zhang and Cheng}]{xian2015fully}
\bibinfo{author}{Xian, M.}, \bibinfo{author}{Zhang, Y.},
  \bibinfo{author}{Cheng, H.D.}, \bibinfo{year}{2015}.
\newblock \bibinfo{title}{Fully automatic segmentation of breast ultrasound
  images based on breast characteristics in space and frequency domains}.
\newblock \bibinfo{journal}{Pattern Recognition} \bibinfo{volume}{48},
  \bibinfo{pages}{485--497}.
\bibitem[{Xiao et~al.(2002)Xiao, Brady, Noble and Zhang}]{xiao2002segmentation}
\bibinfo{author}{Xiao, G.}, \bibinfo{author}{Brady, M.},
  \bibinfo{author}{Noble, J.A.}, \bibinfo{author}{Zhang, Y.},
  \bibinfo{year}{2002}.
\newblock \bibinfo{title}{Segmentation of ultrasound b-mode images with
  intensity inhomogeneity correction}.
\newblock \bibinfo{journal}{IEEE Transactions on Medical Imaging}
  \bibinfo{volume}{21}, \bibinfo{pages}{48--57}.
\bibitem[{Xu et~al.(2019)Xu, Wang, Yuan, Cheng, Wang and
  Carson}]{xu2019medical}
\bibinfo{author}{Xu, Y.}, \bibinfo{author}{Wang, Y.}, \bibinfo{author}{Yuan,
  J.}, \bibinfo{author}{Cheng, Q.}, \bibinfo{author}{Wang, X.},
  \bibinfo{author}{Carson, P.L.}, \bibinfo{year}{2019}.
\newblock \bibinfo{title}{Medical breast ultrasound image segmentation by
  machine learning}.
\newblock \bibinfo{journal}{Ultrasonics} \bibinfo{volume}{91},
  \bibinfo{pages}{1--9}.
\bibitem[{Yang et~al.(2017)Yang, Yu, Wu, Wang, Ni, Qin and Heng}]{yang2017fine}
\bibinfo{author}{Yang, X.}, \bibinfo{author}{Yu, L.}, \bibinfo{author}{Wu, L.},
  \bibinfo{author}{Wang, Y.}, \bibinfo{author}{Ni, D.}, \bibinfo{author}{Qin,
  J.}, \bibinfo{author}{Heng, P.A.}, \bibinfo{year}{2017}.
\newblock \bibinfo{title}{Fine-grained recurrent neural networks for automatic
  prostate segmentation in ultrasound images}, in:
  \bibinfo{booktitle}{Thirty-First AAAI Conference on Artificial Intelligence}.
\bibitem[{Yap et~al.(2017)Yap, Pons, Mart{\'\i}, Ganau, Sent{\'\i}s,
  Zwiggelaar, Davison and Mart{\'\i}}]{yap2017automated}
\bibinfo{author}{Yap, M.H.}, \bibinfo{author}{Pons, G.},
  \bibinfo{author}{Mart{\'\i}, J.}, \bibinfo{author}{Ganau, S.},
  \bibinfo{author}{Sent{\'\i}s, M.}, \bibinfo{author}{Zwiggelaar, R.},
  \bibinfo{author}{Davison, A.K.}, \bibinfo{author}{Mart{\'\i}, R.},
  \bibinfo{year}{2017}.
\newblock \bibinfo{title}{Automated breast ultrasound lesions detection using
  convolutional neural networks}.
\newblock \bibinfo{journal}{IEEE Journal of Biomedical and Health Informatics}
  \bibinfo{volume}{22}, \bibinfo{pages}{1218--1226}.
\bibitem[{Yezzi et~al.(1997)Yezzi, Kichenassamy, Kumar, Olver and
  Tannenbaum}]{yezzi1997geometric}
\bibinfo{author}{Yezzi, A.}, \bibinfo{author}{Kichenassamy, S.},
  \bibinfo{author}{Kumar, A.}, \bibinfo{author}{Olver, P.},
  \bibinfo{author}{Tannenbaum, A.}, \bibinfo{year}{1997}.
\newblock \bibinfo{title}{A geometric snake model for segmentation of medical
  imagery}.
\newblock \bibinfo{journal}{IEEE Transactions on Medical Imaging}
  \bibinfo{volume}{16}, \bibinfo{pages}{199--209}.
\bibitem[{Yu et~al.(2016)Yu, Chen, Dou, Qin and Heng}]{yu2016automated}
\bibinfo{author}{Yu, L.}, \bibinfo{author}{Chen, H.}, \bibinfo{author}{Dou,
  Q.}, \bibinfo{author}{Qin, J.}, \bibinfo{author}{Heng, P.A.},
  \bibinfo{year}{2016}.
\newblock \bibinfo{title}{Automated melanoma recognition in dermoscopy images
  via very deep residual networks}.
\newblock \bibinfo{journal}{IEEE Transactions on Medical Imaging}
  \bibinfo{volume}{36}, \bibinfo{pages}{994--1004}.
\bibitem[{Yu et~al.(2018)Yu, Jiang, Zhou, Qin, Ni, Chen, Lei and
  Wang}]{yu2018melanoma}
\bibinfo{author}{Yu, Z.}, \bibinfo{author}{Jiang, X.}, \bibinfo{author}{Zhou,
  F.}, \bibinfo{author}{Qin, J.}, \bibinfo{author}{Ni, D.},
  \bibinfo{author}{Chen, S.}, \bibinfo{author}{Lei, B.}, \bibinfo{author}{Wang,
  T.}, \bibinfo{year}{2018}.
\newblock \bibinfo{title}{Melanoma recognition in dermoscopy images via
  aggregated deep convolutional features}.
\newblock \bibinfo{journal}{IEEE Transactions on Biomedical Engineering}
  \bibinfo{volume}{66}, \bibinfo{pages}{1006--1016}.
\bibitem[{Yu et~al.(2017)Yu, Tan, Ni, Qin, Chen, Li, Lei and Wang}]{yu2017deep}
\bibinfo{author}{Yu, Z.}, \bibinfo{author}{Tan, E.L.}, \bibinfo{author}{Ni,
  D.}, \bibinfo{author}{Qin, J.}, \bibinfo{author}{Chen, S.},
  \bibinfo{author}{Li, S.}, \bibinfo{author}{Lei, B.}, \bibinfo{author}{Wang,
  T.}, \bibinfo{year}{2017}.
\newblock \bibinfo{title}{A deep convolutional neural network-based framework
  for automatic fetal facial standard plane recognition}.
\newblock \bibinfo{journal}{IEEE Journal of Biomedical and Health Informatics}
  \bibinfo{volume}{22}, \bibinfo{pages}{874--885}.
\bibitem[{Zhang et~al.(2018)Zhang, Dana, Shi, Zhang, Wang, Tyagi and
  Agrawal}]{zhang2018context}
\bibinfo{author}{Zhang, H.}, \bibinfo{author}{Dana, K.}, \bibinfo{author}{Shi,
  J.}, \bibinfo{author}{Zhang, Z.}, \bibinfo{author}{Wang, X.},
  \bibinfo{author}{Tyagi, A.}, \bibinfo{author}{Agrawal, A.},
  \bibinfo{year}{2018}.
\newblock \bibinfo{title}{Context encoding for semantic segmentation}, in:
  \bibinfo{booktitle}{Proceedings of the IEEE conference on computer vision and
  pattern recognition}, pp. \bibinfo{pages}{7151--7160}.
\bibitem[{Zhang et~al.(2017)Zhang, Xie, Xing, McGough and
  Yang}]{zhang2017mdnet}
\bibinfo{author}{Zhang, Z.}, \bibinfo{author}{Xie, Y.}, \bibinfo{author}{Xing,
  F.}, \bibinfo{author}{McGough, M.}, \bibinfo{author}{Yang, L.},
  \bibinfo{year}{2017}.
\newblock \bibinfo{title}{{MDNet}: A semantically and visually interpretable
  medical image diagnosis network}, in: \bibinfo{booktitle}{Proceedings of the
  IEEE conference on computer vision and pattern recognition}, pp.
  \bibinfo{pages}{6428--6436}.
\bibitem[{Zhao et~al.(2017)Zhao, Shi, Qi, Wang and Jia}]{zhao2017pyramid}
\bibinfo{author}{Zhao, H.}, \bibinfo{author}{Shi, J.}, \bibinfo{author}{Qi,
  X.}, \bibinfo{author}{Wang, X.}, \bibinfo{author}{Jia, J.},
  \bibinfo{year}{2017}.
\newblock \bibinfo{title}{Pyramid scene parsing network}, in:
  \bibinfo{booktitle}{Proceedings of the IEEE conference on computer vision and
  pattern recognition}, pp. \bibinfo{pages}{2881--2890}.
\bibitem[{Zhou et~al.(2018)Zhou, Siddiquee, Tajbakhsh and
  Liang}]{zhou2018unet++}
\bibinfo{author}{Zhou, Z.}, \bibinfo{author}{Siddiquee, M.M.R.},
  \bibinfo{author}{Tajbakhsh, N.}, \bibinfo{author}{Liang, J.},
  \bibinfo{year}{2018}.
\newblock \bibinfo{title}{{UNet++}: A nested {U-Net} architecture for medical
  image segmentation}, in: \bibinfo{booktitle}{Deep Learning in Medical Image
  Analysis and Multimodal Learning for Clinical Decision Support}.
  \bibinfo{publisher}{Springer}, pp. \bibinfo{pages}{3--11}.

\end{thebibliography}
\end{document}